\documentclass[prodmode,acmtompecs]{acmsmall} 

\usepackage{gensymb}
\usepackage{cite}
\usepackage{graphicx}
\usepackage{url}
\usepackage{amsmath}
\usepackage{latexsym,amssymb,epsfig,color,verbatim}
\usepackage{algorithm}
\usepackage{algorithmic}
\usepackage[tight,footnotesize,center]{subfigure}
\usepackage{color}
\usepackage{multirow}

\usepackage{wasysym}
\usepackage{xcolor}
\usepackage{tabu}

\newcommand{\BEQA}{\begin{eqnarray}}
\newcommand{\EEQA}{\end{eqnarray}}

\newcommand{\R}{\mathbb{R}}

\newcommand{\X}{\mathbb{X}}

\newcommand{\pr}{\mathbb{P}}
\newcommand{\E}{\mathbb{E}}

\newcommand{\A}{\mathcal{A}}

\newcommand{\ka}{\kappa}

\begin{document}

\markboth{Li et al.}{Mean Field Games in Nudge Systems for Societal Networks}

\title{Mean Field Games in Nudge Systems for Societal Networks}
\author{Jian Li
\affil{Texas A\&M University}
Bainan Xia
\affil{Texas A\&M University}
Xinbo Geng
\affil{Texas A\&M University}
Hao Ming
\affil{Texas A\&M University}
Srinivas Shakkottai
\affil{Texas A\&M University}
Vijay Subramanian
\affil{University of Michigan}
Le Xie
\affil{Texas A\&M University}}

\begin{abstract}
We consider the general problem of resource sharing in societal networks, consisting of interconnected communication, transportation, energy and other networks important to the functioning of society.  Participants in such network need to take decisions daily, both on the quantity of resources to use as well as the periods of usage.  With this in mind, we discuss the problem of incentivizing users to behave in such a way that society as a whole benefits. In order to perceive societal level impact, such incentives may take the form of rewarding users with lottery tickets based on good behavior, and periodically conducting a lottery to translate these tickets into real rewards. We will pose the user decision problem as a mean field game (MFG), and the incentives question as one of trying to select a good mean field equilibrium (MFE).  In such a framework, each agent (a participant in the societal network) takes a decision based on an assumed distribution of actions of his/her competitors, and the incentives provided by the social planner. The system is said to be at MFE if the agent's action is a sample drawn from the assumed distribution.  We will show the existence of such an MFE under general settings, and also illustrate how to choose an attractive equilibrium using as an example demand-response in the (smart) electricity network.
\end{abstract}


\maketitle

\section{Introduction}\label{sec:intro}
There has recently been much interest in understanding \emph{societal networks,} consisting of interconnected communication, transportation, energy and other networks that are important to the functioning of human society.   These systems usually have a shared resource component, and where the participants have to periodically take decisions on when and how much to utilize such resources, but with indirect knowledge of the aggregate utilization of the shared resource.   Research into these networks often takes the form of behavioral studies on decision making by the participants, and whether it is possible to provide incentives to modify their behavior in such a way that the society as a whole benefits \cite{MerPra09,Pra13}.

Our candidate application in this paper is that of a Load Serving Entity (LSE) or a Load Aggregator (LA) (\emph{e.g.,} a utility company) trying to reduce its exposure to daily electricity market volatility by incentivizing demand response in a Smart Grid setting.  The reason for our choice is the ready availability of data and reliable models for the cost and payoff structure that enables a realistic study.  The data used in this paper was obtained from the Electric Reliability Council of Texas \cite{ercot}, an organization that manages the wholesale electricity market in the state.   The price shows considerable variation, and peaks at about $5$ PM each day, which is when maximum demand occurs.   A major source of this demand in Texas is air conditioning, which in each home is of the order of $30$ kWh per day \cite{pecanstreet}. Incentivizing customers to move a few kWh of peak-time usage to the sides of the peak each day could lead to much reduced risks of peak price borne by the LSE.  Such demand shaping could also have a positive effect on environmental impact of power plant emissions, since supplying peak load is associated with inefficient electricity generation.

As an example, we take the baseline temperature setpoint as $22.5\degree$C, and consider a customer that every day increases the setpoint by $1\degree$C in $5-6$ PM and decreases the setpoint by $0.5\degree$C in the off-peak times.   We will see later that even such a small change of the setpoint of the AC, the incentives to the users (a group of fifty homes) can be tuned to achieve different trade-offs: either just a utility increase for the users, just savings to the LSE (of the order of eighty dollars in our case) or any objective that chooses some appropriate mix of the two.  This result is under the implicit assumption that the LSE in question is a price-taker so that changes in its demand profile are assumed not to perturb the prices.  The shifting of daily energy usage could potentially cause a small increase in the mean and deviation of the internal home temperature, which is a discomfort cost borne by the customer.  In our approach, the LSE awards a number of ``Energy Coupons'' to the customer in proportion to his usage at the non-peak times, and these coupons are used as tickets for a lottery conducted by the LSE.   A higher number of coupons would be obtained by choosing an option that potentially entails more discomfort, and would also imply a higher probability of winning at the lottery.  Since the customers do not observe the variation of day-ahead prices on a day-to-day basis nor do they see the aggregate demand at the LSE, the lottery scheme serves as a light-weight and easy to implement mechanism to transfer some of this information over to the customers by coupling them.  We will explicitly demonstrate the advantage of this coupling over an individual incentive scheme (a fixed reward for peak time reductions) that serves as a benchmark for the comparison.

In our analytical model, each agent has a set of actions that it can take in each play of a repeated game, with each action having a corresponding cost.  Higher cost actions yield a higher number of coupons.  Agents participate in a lottery in which they are randomly permuted into groups, and one or more prizes are given in each group.  The state of each agent is measured using his surplus, which captures the history of plays experienced by the agent, and is a proxy to capture his interest in  participating in the incentive system.  A win at the lottery increases the surplus, and a loss decreases it.  Furthermore, we assume that the agent has a prospect incremental utility function that is increasing and concave for positive surplus and convex for negative surplus.  This prospect theory model captures decision making under risk and uncertainty for agents.  Any agent could depart from the system with a  fixed probability independent of the others, and a departing agent is replaced by a new entrant with a randomly drawn surplus.  The main question we answer in this paper then is how would agents decide on what action to take at each play?  Having answered this we also comment on impact of this on the sum total value of the agents, the return to the system and the trade-off between these two quantities provided by our proposed scheme.

\subsection{Prospect Theory}
Most previous studies account for uncertainty in agent payoffs by means of \emph{expected utility theory} (EUT). Here, the objective of the decision maker is to maximize the probabilistically weighted average utilities under different outcomes, and it is assumed that he/she is capable of making arbitrarily complex deductions.  However, EUT does not incorporate observed behavior of human agents, who exhibit bounded rationality and can take decisions deviating from the conventional rational agent norm.  For example, empirical studies have shown that agents ascribe high weights to rare, positive events (such as winning a lottery) \cite{kahneman1979prospect}.

\emph{Prospect theory} (PT) \cite{kahneman1979prospect,tversky1992advances,tversky1981framing,kahneman1984choices} is perhaps the most well-known alternative theory to EUT.  It was originally developed for binary lotteries \cite{kahneman1979prospect} and later refined to deal with issues related to multiple outcomes and valuations \cite{tversky1992advances}.
This Nobel-prize-in-economics-winning theory has been observed to provide a more accurate description of decision making under risk and uncertainty than EUT.  There are three key characteristics of PT. First, the value function is concave for gains, convex for losses, and steeper for losses than for gains. This feature is due to the observation that most (human) decision makers prefer avoiding losses to achieving gains.  Thus, the value function is usually S-shaped.  Second, a nonlinear transformation of the probability scale is in effect, \textit{i.e.}, (human) decision makers will overweight low probability events and underweight high probability events. The weighting function usually has an inverted S-shape, \textit{i.e.}, it is steepest near endpoints and shallower in the middle of the range, which captures the behaviors related to risk seeking and risk aversion. Finally the third, the framing effect is accounted for, \textit{i.e.}, the (human) decision maker takes into account the relative gains or losses with respect to a reference point rather than the final asset position.  As PT fits better in reality than EUT based on many empirical studies, it has been widely used in many contexts such as social sciences \cite{gao2010adaptive,harrison2009expected}, communication networks \cite{li2012prospects,clark2002tussle,yu2014spectrum} and smart grids \cite{wang2014integrating,xiaoprospect}.  Since we study equilibria that arise through (human) agents' repeated play in lotteries, we use PT as opposed to EUT to account for agent-perceived value while taking decisions.

\subsection{Mean Field Games}

The problem described is an example of a dynamic Bayesian game with incomplete information, wherein each player has to estimate the actions of all his potential opponents in the current lottery (and in the future) without knowing their surpluses, play a best response, and update his beliefs about their states of surplus based on the outcome of the lottery.  However, since the set of agents is large and, from the perspective of each agent, each lottery is conducted with a randomly drawn finite set of opponents, an accurate approximation for any agent is to assume that the states of his opponents (and hence actions) are independent of each other.  This is the setting of a Mean Field Game (MFG) \cite{LasLio07,Boyan88,huang2006large}, which we will use as a framework to study equilibria in societal networks.  Here, the system is viewed from the perspective of a single agent, who assumes that each opponent's action would be drawn independently from an assumed distribution, and plays a best response action.   We say that the system is at a Mean Field Equilibrium (MFE) if this best response action turns out to be a sample drawn from the assumed distribution.

\subsection{Demand Response in Deregulated Markets}

Demand Response is the term used to refer to the idea of customers being incentivized in some manner to change their normal electricity usage patterns in response to peaks in the wholesale price of electric power \cite{AlbSaa08}.  Many methods of achieving demand response exist, including an extreme one of turning off power for short intervals to customers a few times a year if the price is very high.  Customers expect a subsidy in return, often in terms of a reduced electricity bill.

\subsection{Main Results}
Our objective in this paper is to design and analyze a system that can provide greater ability for the LSE to realize a desired combination of profit and user value by incentivizing user behavior.  Our main contributions in this paper are as follows:\\ 
(1) We propose a mean field model to capture the dynamics in societal networks. Our model is well suited to large scale systems in which any given subset of agents interact only rarely. This kind of system satisfies a chaos hypothesis that enables us to use the mean field approximation to accurately model agent interactions.  The state of the mean field agent is its surplus, and the agent must choose from a finite set of actions based on its surplus and its belief about the action distribution of other agents.  The state (surplus) evolves according to a Markov process that increases by winning and decreases by losing at the lottery.  Our mean field model of societal networks is quite general, and can be applied to different incentive schemes that are currently being proposed in the field of public transportation and communication network usage. \\
(2) {We conduct a simulation analysis of our scheme under an accurate measurement-based model of the daily usage of electricity in each hour in Texas. 
We also use the data on wholesale electricity prices during the interval to calculate what times of day would yield the best returns to rewards.  {We show that under several intuitive coupon allocation options and a $\$15$ weekly reward (lottery prize), customers would change their AC setpoints (as small as $1\degree$C each day) and each week, the LSE gains a benefit of the order of a $\$80$ over a cluster of $50$ homes.  While doing so, we also numerically verify that our model satisfies conditions needed for passage to the mean field}.}\\
(3) {We conduct comparative studies between a benchmark scheme that returns a fixed reward per action (assuming that each customer maximizes his return) versus the lottery scheme, and show that the lottery scheme can outperform the fixed reward scheme by about $100\%$ in terms of total value to the users, and about $20\%$ in terms of profit to the LSE.  We also explore the relation between LSE profit and user value for both schemes, and show that as one changes the reward values and coupon allocations, the lottery scheme bounds the achievable region of the benchmark scheme in a Pareto-sense: it is better able to attain a desired combination of user value and LSE profit, and includes combinations unachievable by the individual incentive scheme.}\\
(4) We develop a characterization of a lottery in which multiple rewards can be distributed, but with each participant getting at most one by withdrawing the winner in each round.  Each lottery is played amongst a cluster of $M$ agents drawn from a random permutation of the set of all agents.  While the exact form of the lottery is not critical to our results, we present it for completeness. \\
(5) We characterize the best response policy of the mean field agent, using a dynamic programming formulation.  We find that under our assumptions, the value function is continuous in the action distribution, but that multiple actions could turn out to be best responses.  Hence, an agent also needs to choose some randomization method across such equal-value actions.  If the value function is super-modular, sub-modular or S-shaped (under the prospect-based utility function), the action choices map to surplus intervals, with two actions being of equal value at each interval boundary.\\
(6) The probability of winning the lottery defines the transition kernel (along with the regeneration distribution) of the Markov process of the surplus, and hence maps an assumed distribution across competitors states to a resultant stationary distribution.  We show the existence of a fixed point of this kernel, which is the MFE, by using Kakutani's fixed point theorem.  Essentially, the system is a map between the space consisting of the triple of an assumed action distribution, a randomized policy and a surplus distribution back to itself, and our result is to show this map has a fixed point.  Our proof of the existence of MFE does not depend on the shape of the utility function, which can be quite general.  Since we have a discrete action and state space, showing a fixed point in the space of such triples is quite intricate.

A 2-page conference abstract that includes a high-level overview of our results developed herein was presented to practitioners in  \cite{JianBai15}.

\subsection{Related Work}
In terms of the MFG, our framework is based on work such as  \cite{IyeJoh14,ManRam14,JianRaj16ton}.  In \cite{IyeJoh14} the setting is that of advertisers bidding for spots on a webpage, and the focus is on learning the value of winning (making a sale though the advertisement) as time proceeds.  In \cite{ManRam14}, apps on smart phones bid for service from a cellular base station, and the goal is to ensure that the service regime that results has low per-packet delays.  In both works, the existence of an MFE with desired properties is proved.   In \cite{JianRaj16ton}, the objective is to incentivize truthful revelation of state that would allow for optimal resource allocation in a device-to-device wireless network.  The state space is discrete, and the focus is on the exploration of truthful dynamic mechanisms in the mean field regime.   However, unlike that work, we focus on a lottery-based allocation in this paper.  The lottery is simple and well-established, and has been successfully applied in a variety of existing nudge systems.  Thus, our goal is to analyze this well-established mechanism, rather than designing new ones.  Also, unlike the previous work, all of which focussed on pure strategy equilibria, our current work has a more complex state space and pure strategy equilibria may not exist due to the non-uniqueness of best responses.  Hence, we seek a mixed strategy equilibrium, which necessitates a different proof technique.

Nudge systems are typically designed and used to encourage socially beneficial behaviors and individually beneficial behaviors.  For instance, lottery schemes are widely used in practice to incentivize good behavior, e.g., to combat (sales) tax evasion in Brazil (\cite{jnari13}), Portugal (\cite{polo15}), Taiwan (\cite{chen2010tax}),  and for Internet congestion management (\cite{patrick14}).  Similarly, \cite{MerPra09,Pra13} provide experimental results on designing lottery-based ``nudge engines'' to provide incentives to participants to modify their behaviors in the context of evenly distributing load on public transportation.  In another scheme, \cite{allcott2015welfare} study the impact of nudging on social welfare by sending one-year home energy reports to participants and using multiple price lists to determine participants' willingness to stay in the system for the next year.  Our system is a form of nudge engine, but our focus is on analytical characterization of system behavior and attained equilibria with large number of customers with repeated decision-making. We aim to design incentive schemes to modify customer behavior such that the system as the whole benefits from the attained equilibrium.

Our idea of offering coupons for reduced electricity usage at certain times is based on one presented in \cite{ZhoXie13}, which suggests offering incentives to coincide with predicted realtime price peaks.  An experimental trial based on a similar idea is described in \cite{bitar15}, in which the focus is on designing algorithms to coordinate demand flexibility to enable the full utilization of variable renewable generation.  In \cite{HaoXie14}, this kind of system is modeled as a Stackelberg game with two stages: setting the coupon values followed by consumer choice. The decision making model in all the above research is myopic.  The authors of \cite{galina12} study demand-response as trading off the cost of an action (such as modifying energy usage) against the probability of winning at a lottery in terms of a mean field game.  However, the game is played in a single step according to their model, and there is no evolution of state or dynamics based on repeated play.  Further, their conception of the mean field equilibrium is that the mean value of the action distribution (not the distribution itself) is invariant.  Unlike these models, we are interested in characterizing repeated consumer choice with state evolution when the number of customers is large, and identifying the action distribution and benefits (if any) of the resulting equilibrium.

A rich literature studies lottery schemes, and here we can only hope to cover a fraction of them that we see most relevant. In this paper, we model lotteries as choosing a random permutation of the $M$ agents participating in it, and picking the first $K$ of them as winners, with the distribution on the symmetric group of permutations of $\{1, \cdots, M\}$ being a function of the coupons assigned to the different actions. Assuming that different actions yield different numbers of coupons, we will choose the distribution such that more coupons results in a higher probability of winning. There are various probabilistic models on permutations in the ranking literature \cite{qin2010new,lozano12},  Here we use the popular Plackett-Luce model \cite{hunter2004mm} to implement our lotteries. While the Plackett-Luce model is used for concreteness, other probabilistic models on permutations such as the Thurstone model \cite{lozano12} can also be used with the number of coupons as parameters of the distribution as long as more coupons results in a higher probability of winning.

The monotonicity properties in rewards are shared with other literature, such as \cite{gomes10,adlakha15}.  In particular, \cite{adlakha15} focuses on the existence of the mean field equilibrium when players¡¯ welfare depends on the distribution of other players actions.   However, this previous work studies the existence of pure strategy equilibria, whereas our discrete state and action spaces requires consideration of a mixed strategy equilibrium. The proof of the existence of this equilibrium is one of the major technical contributions of this paper.

\subsection{Organization}
The paper is organized as follows.  In Section~\ref{sec:mfe-model}, we introduce our mean field model.  {We then conduct simulation-based numerical studies in Section~\ref{sec:numerical}, on utilizing our framework in the context of demand response in electricity markets.}  In Section~\ref{sec:lottery} we develop a characterization of a lottery in which multiple rewards can be distributed, but with each participant getting at most one by withdrawing the winner in each round. We discuss the basic property of the optimal value function in Section~\ref{sec:opt-val-fn}. The existence of MFE is considered in Section~\ref{sec:mfe-definition}. We characterize the best response policy of the mean field agent, using a dynamic programming formulation in Section \ref{sec:best-resp}. We conclude in Section~\ref{sec:conclusion}. To ease exposition of our results, all proofs are relegated to the Appendix.

\section{Mean Field Model}\label{sec:mfe-model}
We consider a general model of a societal network in which the number of agents is large.  Agents have a discrete set of actions available to them, and must take one of these actions at each discrete time instant.  The actions result in the agents receiving coupons, with higher cost actions resulting in more coupons.  The agents are then randomly permuted into clusters of size $M,$ and a nudge is provided via a lottery that is held using the coupons to win real rewards.  Thus, agents must take their actions under some belief about the likely actions, and hence the likely coupons held by their competitors in the lottery.  

\begin{figure}[ht]
\begin {center}
\includegraphics[width=3.3in]{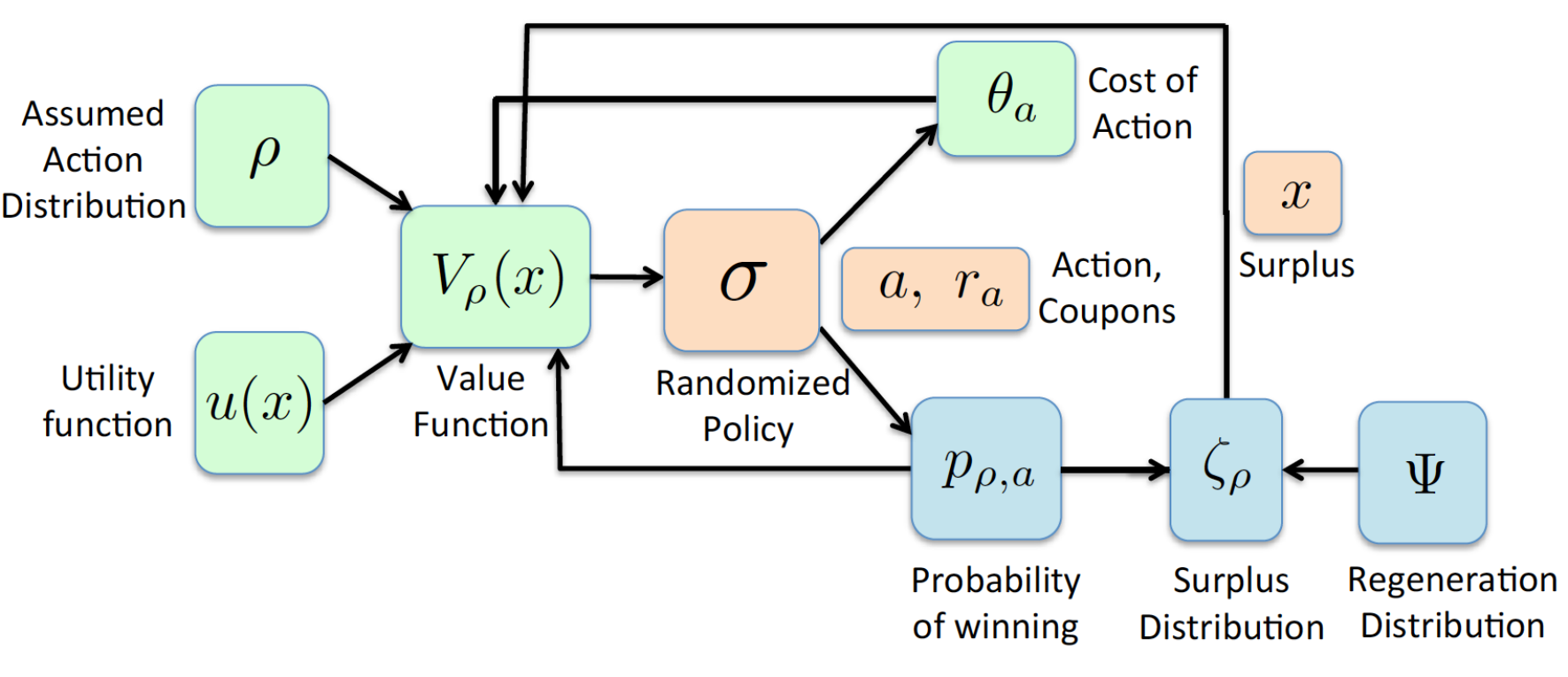}
\caption{Mean Field Game.}
\label{fig:MFE-diagram}
\vspace{-0.2in}
\end{center}
\end{figure}

Figure~\ref{fig:MFE-diagram} illustrates the mean field approximation of our model.  We provide justification for the mean field approximation in the discussion at the end of this section.   The diagram is drawn from the perspective of a single agent  (w.l.o.g, let this be agent $1$), who assumes that the actions played by each of his opponents would be drawn independently of each other from the probability mass function $\rho.$   In this section, we will introduce the notation, costs and payoffs of the agent, and provide a brief description of the policy space and equilibrium.  

\noindent{\bf{Time}:} Time is discrete and indexed by $k\in\{0, 1, \cdots\}$.

\noindent{\bf{Agents}:} The total number of agents is infinite, and we consider a generic agent $1$ who in each lottery will be paired with $M-1$ others drawn randomly.

\noindent{\bf{Actions}:} We suppose that each agent has the same action space denoted as $\A=\{1, 2,\cdots, |\mathcal{A}|\}.$ Hence, the action that this agent takes at time $k$ is $a[k]\in \A$.  Under the mean field assumption, the actions of the other agents would be drawn independently from the p.m.f. $\rho =[b_1,b_2,\cdots, b_{|\mathcal{A}|}],$ where $b_a$ is the probability mass associated with action $a.$  We call $\rho$ as the assumed action distribution.

\noindent{\bf{Costs}:} Each action $a\in \A$ taken at time $k$  has a corresponding cost $\theta_a.$  This cost is fixed and represents the discomfort suffered by the agent in having to take that action.

\noindent{\bf{Coupons}:} When agent takes an action $a$, it is awarded some fixed number of coupons $r_a$ for playing that action.  These coupons are then used by the agents as lottery tickets.

\noindent{\bf{Lottery}:} We suppose that there are only $K$ rewards for agents in one cluster, where $K$ is a fixed number less than $M$.  The probability of winning is based on the number of coupons that each agent possesses. We model each lottery as choosing a permutation of the $M$ agents participating in it, and picking the first $K$ of them as winners.  {We denote the winning probability as $p_{\rho, a}$ and derive its explicit form in Section~\ref{sec:lottery}.}

\noindent{\bf{States}:} The agent keeps track of his history of wins and losses in the lotteries by means of his net surplus at time $k,$ denoted as $x[k].$  The value of surplus is the state of the agent, and is updated in a Markovian fashion as follows:
\begin{equation}
x[k+1]=
\begin{cases}
\begin{aligned}
&x[k]+w, & \text{if agent $1$ wins the lottery}, \\
&x[k]-l, & \text{if agent $1$ loses the lottery},
\end{aligned}
\end{cases}
\end{equation}
where $w$ and $l$ is the impact of winning or losing on surplus.  Effectively, the assumption is that the agent expects to win at least an amount $l$ at each lottery.  Not receiving this amount would decrease his surplus.  Similarly, if the prize money at the lottery is $w+l,$  the increase in surplus due to winning is $w.$   
Surplus values are discrete, and the set of possible values is given by a countable $\X$ that ranges from $(-\infty,+\infty)$. 

\noindent{\bf{Value function for prospect}:}  The impact of surplus on the agent's happiness is modeled by an S-shaped incremental utility function $u(x[k])$, which is monotone increasing, concave for a positive surplus and convex for a negative surplus. Moreover, the impact of loss is usually larger than that of gain of the same absolute value.  Note that we implicitly assume that the reference for all agents is $0$. Then following \cite{tversky1992advances}, we use the following value function for prospect
\begin{equation}
\label{eqn:utility-fn}
\begin{aligned}
u(x)=\begin{cases} 
        u^+(x)=x^{\gamma},  & x\geq 0,\\
        u^-(x)=-{\varphi} (-x)^{\gamma}, & x<0,
        \end{cases}
\end{aligned}
\end{equation} 
where $\varphi>1$ is the loss penalty parameter and $0<\gamma<1$ is the risk aversion parameter. A larger $\varphi$ means that the operator is more loss averse, while a smaller $\gamma$ 
 indicates that the operator is more risk seeking.  From empirical studies \cite{tversky1992advances,kahneman1984choices}, realistic values are $\varphi=2.25$ and $\gamma=0.88.$

\noindent{\bf{Weighting function for prospect}:} It has been observed empirically that people tend to subjectively weight uncertain outcomes in real-life decision making \cite{prelec1998probability}.  In the proposed game, this weighting factors capture the agent's subjective evaluation on the mixed strategy of its opponents.  Thus, under PT, instead of objectively observing the probability of winning the lottery $p_{\rho, a}$, each user perceives a weighted version of it, $\phi(p_{\rho,a})$.  Here, $\phi(\cdot)$ is a nonlinear transformation that maps the objective probability to a subjective one, which is monotonic increasing in probability. It has been shown in many PT studies that, people usually overweight low probability outcomes and underweight high probability outcomes.
Following \cite{prelec1998probability}, we use the weighting function
\begin{equation}
\phi(p)=\exp(-(-\ln p)^{\xi}), \quad \text{for}\quad 0\leq p\leq 1,
\label{eq:weighting}
\end{equation}
where $\xi\in (0,1]$ is the objective weight that characterizes the distortion between subjective and objective probability.  Note that under the extreme case of $\xi=1$,~(\ref{eq:weighting}) reduces to the conventional EUT probability, i.e., $\phi(p)=p$.

\noindent{\bf{Regeneration}:} An agent may quit the system at any time, independent of others.  This event occurs with probability $1-\beta,$ where $\beta \in (0,1).$  When this happens, a new agent takes the place of the old one, with a state drawn from a probability mass function $\Psi.$

\noindent{\bf{Best Response Policy}:}  The agent must choose an action at each time, including staying with the status-quo/baseline as an action too. The green/light tiles in Figure~\ref{fig:MFE-diagram} relate to the problem of the agent determining his best response policy.  The agent assumes that the actions taken by  each of his $M-1$ opponents are drawn independently from probability mass function $\rho.$  Given this assumption, the state of his surplus is $x$ and current utility is $u(x),$  the agent must calculate the probability of winning at the lottery $p_{\rho,a}(x),$ if he were to take action $a(x) \in \A,$ incurring a cost $\theta_{a(x)}$ and gaining $r_{a(x)}$ coupons.  Since the agent must take this decision repeatedly, he must solve a dynamic program to determine his optimal policy. There could be many best response actions, and we assume that the agent chooses a randomized policy $\sigma(x)\triangleq [\sigma_1(x), \sigma_2(x),\cdots, \sigma_a(x),\cdots, \sigma_{|\mathcal{A}|}(x)],$ in which $\sigma_a(x)$ specifies the probability of playing action $a$ when the agent's surplus is $x;$ in other words, we enlarge the space to include mixed strategies as pure strategy equilibria may not exist.  The action taken by the agent is a random variable $A\sim\sigma(x).$ The details of the lottery and how to calculate the probability of success are given in Section~\ref{sec:lottery}.  The properties of the best response policy are described in detail in Section~\ref{sec:best-resp}.  

\noindent{\bf{Stationary Surplus Distribution}:}  The assumed action distribution $\rho$, and the best-response randomized policy $\sigma(x)$ yield the state transition kernel of the Markov chain corresponding to the surplus, via the probability of winning the lottery $p_{\rho,a}(x)$.  This is illustrated by means of the blue/dark tiles in Figure~\ref{fig:MFE-diagram}.  The transition kernel also is influenced by the regeneration distribution $\Psi.$  The stationary distribution of surplus associated with the transition kernel is denoted as $\zeta_\rho.$    This stationary distribution of the single mean field agent is equivalent to the one-step empirical state distribution of infinite agents who all take a (mixed-strategy) action, $\sigma(x)$ when state is $x$, assuming that the actions of their competitors would be drawn from $\rho.$ 

\noindent{\bf{Mean Field Equilibrium}:}  The triple of an assumed action distribution $\rho$, randomized policy $\sigma$ and stationary surplus distribution $\zeta$ gets mapped via mapping $\Pi^*$ into a triple of action distribution $\tilde{\rho}$, best-response randomized policy $\tilde{\sigma}$ and a stationary surplus distribution $\tilde{\zeta}$ via the operations described above.  A fixed point of the resulting map is called an MFE.  For a formal definition and the proof of existence see  Section~\ref{sec:mfe-definition}.

\subsection{Discussion}\label{sec:mfe-discussion}

Is the MFG a good approximation?  Specifically, we need to first show that for any agent, the assumption that the states of any finite subset of agents that it interacts with are independent of it and each other as the number of agents becomes asymptotically large.  Second, we need to show that when we repeat the game over time, the empirical distribution of the agents' states converges to a fixed point (mean field limit).

The first result is follows from an argument called \emph{propagation of chaos} via constructing \emph{interaction sets} defined in \cite{GraMel94}, which characterize the conditions under which any finite subset of the state of the agents are independent of each other.   Following a similar argument to \cite{IyeJoh14}, we can show that after any finite number of lotteries (finite time), as the total number of agents becomes large enough, the interaction sets of any finite collections of agents become disjoint with high probability.  Hence, the states of these agents become independent.  Inspired by \cite{IyeJoh14}, the proof is divided into two parts: (i) first, we need to show that as the total number of agents $JM$ ($J$ is the number of lotteries) becomes large enough, the probability that agent $1$ interacted with the set of agents (that it interacts at the $k$-th lottery, $k\geq 1$) before the $k$-th lottery become zero; and  (ii) the action distribution $\rho_1$, the randomized policy $\sigma_1,$ and the surplus distribution $\zeta_1$ of agent $1$ converges to the assumed distributions $\rho, \sigma, \zeta,$ respectively, as the number of agents $JM$ becomes large enough.  We do not present the full argument here due to space limitations and the fact that it follows via identical arguments to \cite{GraMel94,IyeJoh14}.

{The second result requires the establishment of the so called \emph{Mckean-Vlasov limit}---a differential equation that specifies the evolution of the empirical distribution of state over the transition kernel specified in Figure~\ref{fig:MFE-diagram}.  In order to do this, we need to verify three sufficiency conditions presented in \cite{borkar13})  (see Section $2$ Assumptions $\boldsymbol{A1}$ -  $\boldsymbol{A3}$) built on a continuous-time Markov chain (CTMC).  It is easy to move our discrete-time Markov chain (DTMC) setup to their framework by equipping each agent with an independent Poisson clock with rate $\lambda_Q$ (chosen as $1$ w.l.o.g).   An agent whose clock ticks is allowed to take an action, and receives a reward with the same probability engendered by a lottery under the same action and with the same belief distribution.  The equivalence of the stationary distributions of the CTMC and DTMC versions follows immediately from \cite{ross2013applied} (Chapter 7), with the Bellman equation of the DTMC system being replaced by the Hamilton-Jacobi-Bellman equation of the CTMC.  In our problem, the Q-matrix of the equivalent CTMC is simply $- \lambda_Q I + \lambda_Q P,$ where $P$ is the P-matrix of the DTMC version and $I$ is the identity matrix.   The most important condition of \cite{borkar13} that needs to be verified is the assumption on the Lipschitz nature of the map between the belief action distribution and the resultant action distribution.  We numerically verify in Section~\ref{sec:numerical-mfe} that given an action belief $\rho,$ the derivative of this map at each iteration step is bounded, leading to the desired Lipschitz property for both DTMC and CTMC.  While this supports the conjecture that the condition holds in our case, the proof is beyond the scope of this work due to the implicit form of the map.}

\section{Numerical Study}\label{sec:numerical}
We conduct an empirical data-based simulation in the context of electricity usage for home air conditioning to illustrate the likely performance of our nudge system in the context of electricity demand-response.  In doing so, we will also numerically study the properties of the mean field approximation.  As mentioned in Section~\ref{sec:intro}, our context is that of a Load Serving Entity (LSE) trying to incentivize its customers to shape their electricity consumption so as to reduce its cost of electricity purchase from the wholesale market whose price variation is as shown in Figure \ref{fig:day-ahead}.  These incentives could increase the net surplus of the end-users, the profit of the LSE, or the total welfare of these agents as well.  Data available for our simulations consist of historical electricity prices from \cite{ercot}, and a data set containing appliance-wise electricity usage for about $1000$ homes along with the ambient temperatures over each day in June--August, 2013 \cite{pecanstreet}.

\subsection{Home Model}
A standard continuous time model \cite{Cal09,Hao14} for describing the evolution of the internal temperature $\tau(t)$ at time $t$ of an air conditioned home is
\begin{equation}\label{hometemp}
\dot{\tau}(t)=
\begin{cases}
\begin{aligned}
&-\frac{1}{RC}(\tau(t)-\tau_a)-\frac{\eta}{C} P_m, &\text{if $q(t)=1$},\\
&-\frac{1}{RC}(\tau(t)-\tau_a),                       &\text{if $q(t)=0$}.
\end{aligned}
\end{cases}
\end{equation}
Here, $\tau_a$ is the ambient temperature (of the external environment), $R$ is the thermal resistance of the home, $C$ is the thermal capacitance of the home, $\eta$ is the efficiency, and $P_m$ is the rated electrical power of the AC unit. The state of the AC is described by the binary signal $q(t),$ where $q(t)=1$ means AC is in the ON state at time $t$ and in the OFF state if $q(t)=0$.   The state is determined by the crossings of user specified temperature thresholds as follows:
\begin{equation}\label{homecontrol}
\lim_{\epsilon \rightarrow 0} q(t +\epsilon)=
\begin{cases}
\begin{aligned}
&q(t),  & |\tau(t) -\tau_r| \leq \Delta,\\
&1,  & \tau(t) >\tau_r+ \Delta,\\
&0,  & \tau(t) <\tau_r-\Delta,
\end{aligned}
\end{cases}
\end{equation}
where $\tau_r$ is the temperature setpoint and $\Delta$ is the temperature deadband.
\begin{table}[ht]
\vspace{-0.1in}
 \tbl{Parameters for a Residential AC Unit}{
\vspace{-0.05in}

\begin{tabular}{|c|c|c|c|c|c|}
\hline
 $C$(Capacitance) & $R$(Resistance) & $P_m$(Power) & $\eta$(Coefficient) & $\tau_r$(Setpoint) & $\Delta$(Deadband) \\ \hline
 $10$ kWh/\degree C & $2$  \degree C/kW & $6.8$ kW & $2.5$ & $22.5$ \degree C & $0.3$ \degree C \\ \hline
\end{tabular}}
 \vspace{-0.1in}
 \label{tbl:parameters}
\end{table}

A number of studies investigate the thermal properties of typical homes.  We use the parameters shown in Table~\ref{tbl:parameters} for our simulations.  These are based on the derivations presented in \cite{Cal09} for temperature conditioning a $250$ m$^2$ home (about $2700$ square feet), which is a common mid-size home in many Texas neighborhoods.

\begin{figure}[ht]
\begin {center}
\begin{minipage}{.3\textwidth}
  \centering
\includegraphics[width=1\columnwidth]{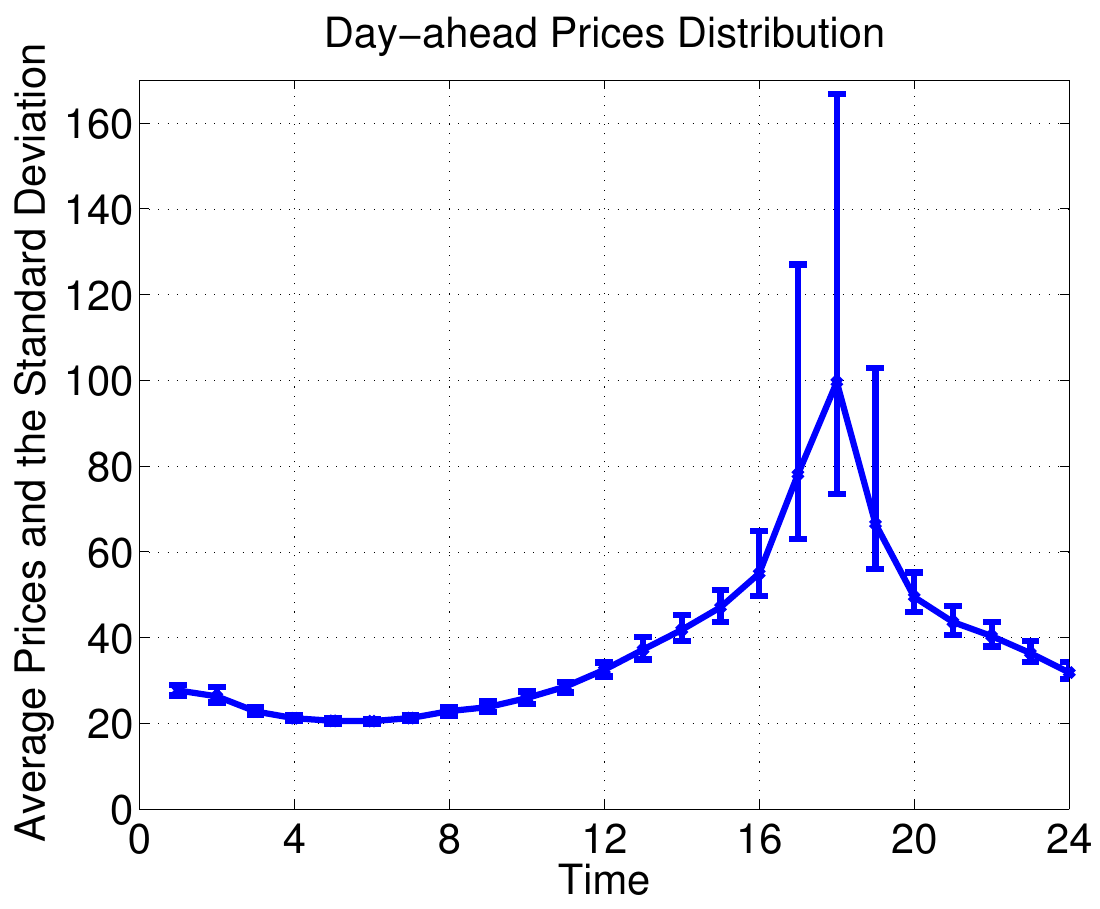}
\caption{Day-ahead electricity market prices in dollars per MWh on an hourly basis between 12 AM to 12 PM, measured between June--August, 2013 in Austin, TX.  Standard deviations above and below the mean are indicated separately.}
\label{fig:day-ahead}
\end{minipage}\hfill
\begin{minipage}{.3\textwidth}
  \centering
\vspace{-0.15in}
\includegraphics[width=1\columnwidth]{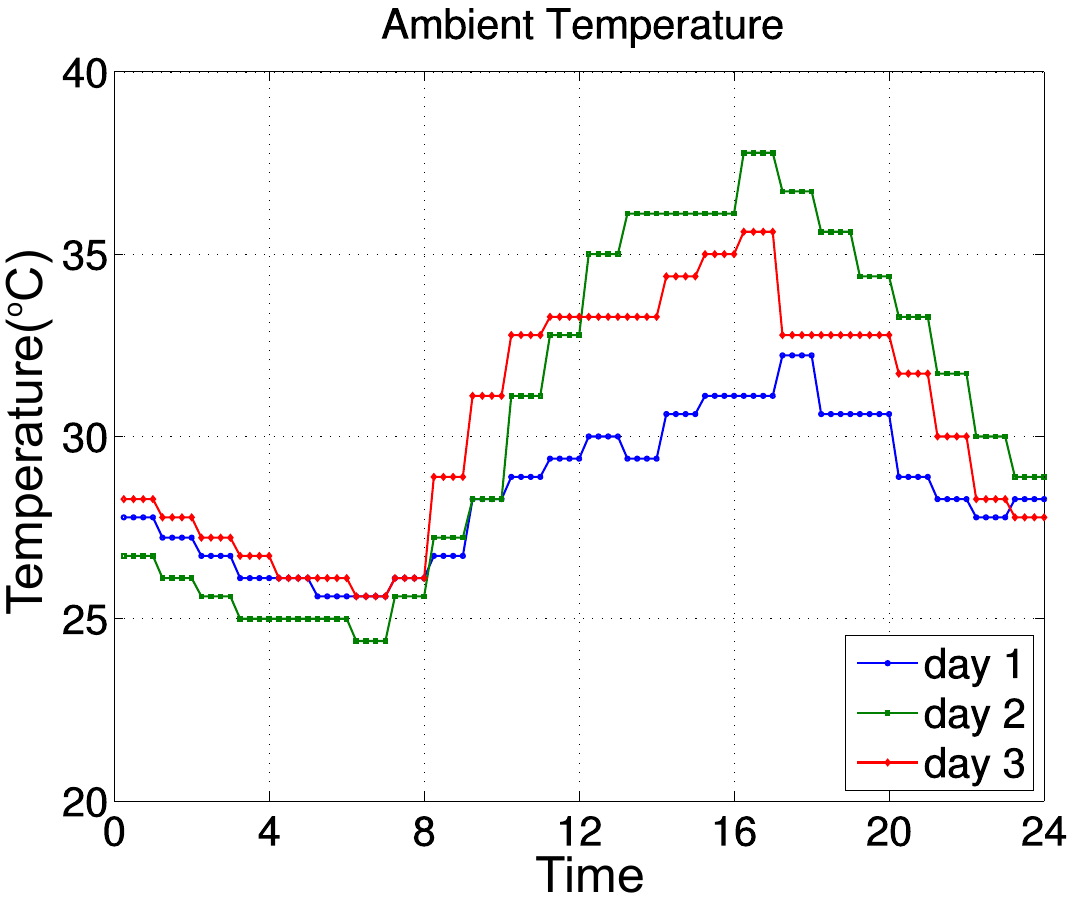}
\caption{Ambient temperature of 3 arbitrary days from June--August, 2013 in Austin, TX. Measurements are taken every $15$ minutes from $12$ AM to $12$ PM.}
\label{fig:ambient}
\end{minipage}\hfill
\begin{minipage}{.3\textwidth}
  \centering
\includegraphics[width=1\columnwidth]{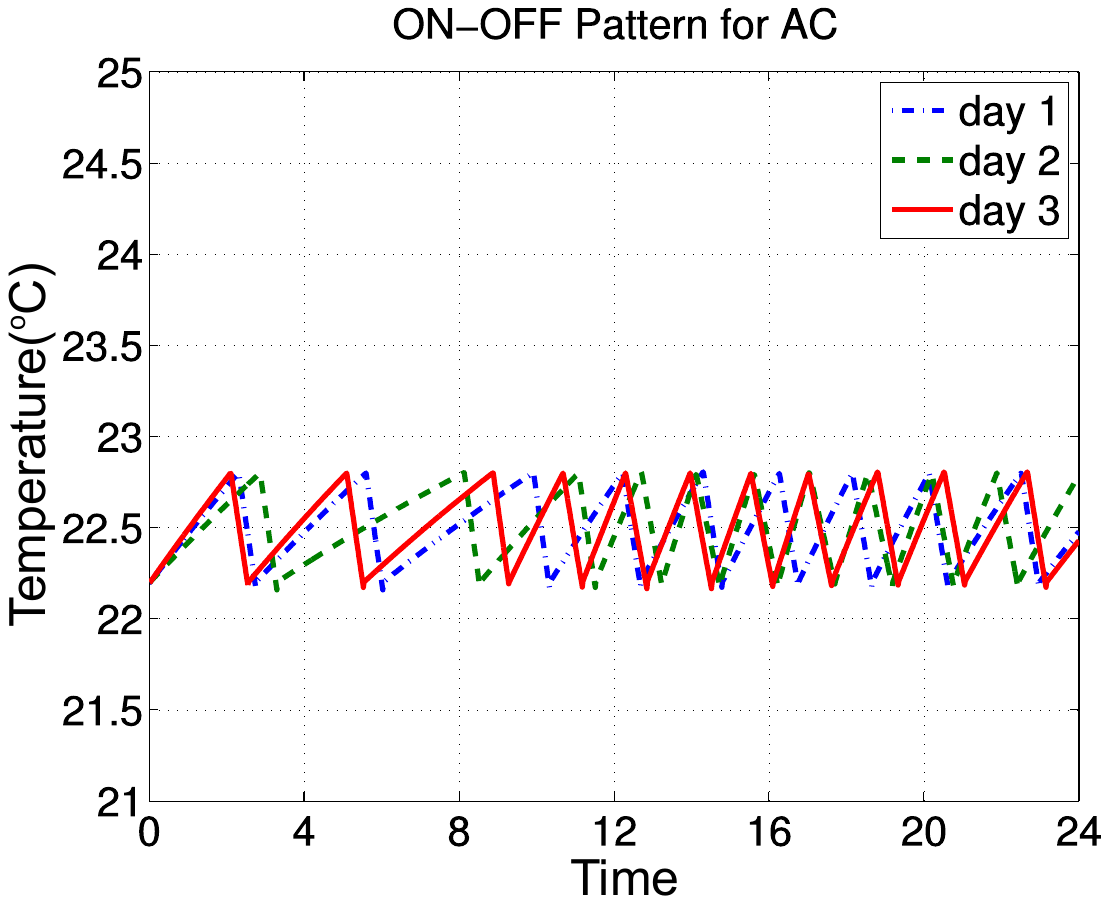}
\caption{Simulated ON/OFF state of AC over a $24$ hour period in a home and the corresponding interior temperature.  The interior temperature falls when the AC comes on, and rises when it is off.}
\label{fig:onoff}
\end{minipage}\hfill
\end{center}
\end{figure}

In order to determine the energy usage for AC in our typical home, we need to know how the ambient temperature varies in Texas during the summer months of interest.  These values are available in the Pecan Street data set,  and we plot the values of $3$ days which are arbitrarily chosen over three months for Austin, TX in Figure~\ref{fig:ambient}.

Next, we calculate the ON-OFF pattern of our typical air conditioner based on the ambient temperature variation over the course of the day.  We do this by simulating the controller in (\ref{homecontrol}) with the appropriate ambient temperature values taken from Figure~\ref{fig:ambient}.  The pattern is presented in Figure~\ref{fig:onoff}. We see that there is higher energy usage during the hotter times of the day, as is to be expected.  This also corresponds to the peak in wholesale electricity prices shown for the same period in Figure~\ref{fig:day-ahead}.  The total energy used each day corresponding to our $2500$ sq ft home with a $5$ ton AC ($=6.8$ kW; see Table~\ref{tbl:parameters})  is $32.83$ kWh.  For comparison, we identified $4$ homes in the Pecan Street data set that have parameters in the same ballpark as our typical (simulated) home.  The average size of these real homes was $2627$ sq feet, with a $4$ ton AC on average, and the average electricity consumed for airconditioning was $34.8$ kWh per day during time interval corresponding to our simulation.  The numbers are quite similar to our simulated home, indicating accuracy of the model.

\subsection{Actions, Costs and LSE Savings}
Since we are interested in peak-period usage, we consider an action set available to the customer that consists of choosing different thermostat setpoints during each hour from $2-8$ PM, i.e, $6$ periods (hours) in total.  We denote each period by an index $j,$ where $j=1$ indicates the period $2-3$ PM and so on until $j=6,$ which indicates the period $7-8$ PM.  Each action can now be identified with a vector $(y_1,y_2,y_3,y_4,y_5, y_6),$ where $y_j$ indicates the setpoint in the period $j.$  We take the setpoint $22.5 \degree$C as the baseline.  Hence, the vector $(22.5,22.5,22.5,22.5,22.5, 22.5)$ indicates a baseline action in which the customer does not change the original setpoint in each period.  The set of all such setpoint vectors defines an action set $\mathcal{A},$ and we define the action with index $a=0$ to be the no-change action.  Since the setpoints on a thermostat are discrete, the number of actions is finite.  We identified $5$ other actions that appeared to have the most promise of being used.  These actions are shown in the second column of Table~\ref{tbl:actions}.

We next calculate the cost of taking each action $a\in\A$, which corresponds to the discomfort of having a potentially higher mean and standard deviation in the home temperature, and possibly higher energy consumption.  We measure the state of the home under action $a\in\mathcal{A}$ by the tuple consisting of the mean temperature, the standard deviation and energy usage, denoted $[\bar{\tau}_a,\sigma_a, \text{E}_a].$  The baseline state of these parameters is under action $0,$ denoted by $[\bar{\tau}_0, \sigma_0, \text{E}_0]$   We define the cost of taking any action $a$ as
\BEQA\label{sim:cost}
\theta_a = |\bar{\tau}_0-\bar{\tau}_a| + \lambda |\sigma_0-\sigma_a|-\varsigma(\text{E}_0-\text{E}_a),
\EEQA
where we choose $\lambda = 10$ to make the numerical values of the mean and standard deviation comparable to each other and $\varsigma=10$ \cent/kWh as the fixed energy price.  We note that the map between temperature variation, discomfort suffered, and its measurement in cents is not obvious.  However, given the fact that the  customer uses between $1-3$ kWh or about $\cent 10-30$ per hour to obtain a temperature differential between the ambient temperature and interior temperature of about $15-20$\degree C, the discomfort cost of a degree C temperature increase being $\cent 1$ seems reasonable in the limited temperature range that we are interested in.  Note that the calculation of cost for each action involves simulating the home under that action to determine $[\bar{\tau}_a,\sigma_a, \text{E}_a].$  However, this has to be done only once to create a look-up table, which can be used thereafter.  Note also that each action in $\mathcal{A}$ is chosen to be close to energy neutral, i.e., the third term in (\ref{sim:cost}) is essentially zero.  Thus, we focus on modifying usage time, not the total usage.  Table~\ref{tbl:actions} shows our selection of actions and their corresponding costs.

When applied over a day, each action could result in some savings to the LSE towards the costs it incurs in purchasing electricity.  We measure the day-ahead price of electricity experienced by the LSE in dollars/MWh and denote the price at time period $j$ in day $i$ as $\pi_{i,j}$, where $i=\{1, 2, \cdots, 92\}$ and $j=\{1, \cdots, 6\}$.
Each action vector of a customer would impose a net price on the LSE in proportion to the usage. We define the \emph{differential price} measured in dollars imposed by an action $y=(y_1,y_2,y_3,y_4,y_5, y_6)$ versus $z=(z_1,z_2,z_3,z_4,z_5, z_6)$  as
\BEQA
H(y,z) =\sum_j \left(k(y_{i,j})-k(z_{i,j})\right)\pi_{i,j},
\EEQA
where $k$ converts the setpoints into electricity usage in each period, which is measured in MWh. Setting $y$ as the baseline action $(22.5,22.5,22.5,22.5,22.5,22.5)$ presents a way of measuring the reduction/increase in cost due to the incentive scheme.

{We calculated the savings of each action applied over each day of our three month data set and obtained the average savings.  These values are shown in  Table~\ref{tbl:actions} (where the final columns entitled ``C$_0$--C$_3$'' will be discussed in Section~\ref{sec:lottery-params}).  As is clear, the cost of taking each of our selected actions is {considerably lower} than the savings resulting from that action, and hence it might be possible to create appropriate incentive schemes to encourage their adoption.  }We will consider two such schemes, namely, (i) a fixed reward scheme used as a benchmark, and (ii) a lottery based scheme.


\begin{table}[ht]
\vspace{-0.1in}
 \tbl{Actions, Costs, LSE Savings and Coupons Awarded}{
\vspace{-0.05in}
 \begin{tabu}{|c||c|c|c|c|c|c|c|}
 \hline
Index	& Action Vector 	& Cost (\cent)	        & LSE    & {C$_0$}   &{C$_1$}& C$_2$&{C$_3$}  \\ 
	& 	& 	        & Savings (\cent)       &  & & &  \\ \hline
$0$ 	               &$(22.5, 22.5, 22.5, 22.5, 22.5, 22.5)$ 	       &$0$              &$0$    &$37.4$       &$8.416$      &$8.416$    &$8.416$       \\
$1$ 	               &$(21.5, 21.5, 22.25, 23.5, 23.75, 21.25)$ 	       &$3.68$  	&$27.7$    &$715$       &$431.8$      &$521.8$     &$611.8$      \\
$2$ 	               &$(21.5, 21.5, 22.25, 23.5, 23.25, 22.25)$ 	        &$3.15$  &$22.7$    &$693$        &$431.8$       &$511.6$     &$591.5$       \\
$3$ 	               &$(21.5, 21.5, 22.25, 24, 23, 22.5)$ 	                &$2.68$  	&$22$     &$577$        &$431.8$       &$466.8$     &$501.7$         \\
$4$ 	               &$(22, 22, 22.25, 23, 23, 22.5)$ 	                         &$1.34$   &$19$     &$434$         &$287.4$       &$325.6$      &$363.7$          \\
$5$ 	               &$(22, 22, 22.25, 23.25, 22.5, 22.75)$ 	        &$0.95$  &$16.4$    &$222$        &$287.4$       &$244.2$      &$200.9$         \\	\hline
 \end{tabu}}
 \vspace{-0.2in}
 \label{tbl:actions}
\end{table}

\subsection{Benchmark Incentive Scheme}\label{sec:benchmark}

A simple incentive scheme to get users to adopt cost saving actions (from the LSE's perspective) is to calculate the expected savings of each action, and to deterministically reward each agent with some percentage of the expected savings for taking that action.  Such guaranteed savings are similar in spirit to rebate for using public transportation during off-peak hours, and a system of sharing a fraction of the savings by demand-response providers such as OhmConnect~\cite{ohm-connect}. We will use this scheme as a benchmark in order to determine whether shared savings are large enough to encourage meaningful participation.  Thus, our benchmark incentive scheme attempts to incentivize each action by returning to the user some fixed fraction of the expected LSE savings for that action presented in Table~\ref{tbl:actions}.  For example, a return of $50\%$ for taking action $1$ would imply awarding $\cent13.85$ each time that action is taken.

\subsection{Lottery-Based Incentive Scheme}\label{sec:lottery-params}
{Our second incentive scheme is lottery-based, with Energy Coupons being used as lottery tickets.  Now, the baseline action $a=0$ corresponds to a setpoint of $22.5 \degree$C in period $3$ at which $\pi_{i,3}$ is highest (Figure~\ref{fig:day-ahead}) for any day $i$.   Hence, the LSE should incentivize actions that are likely to reduce the risks of peak day-ahead price by offering Energy Coupons in proportion to the usage during the corresponding periods.  In the context of our simulation, it is intuitively clear that coupons must be placed at periods of lower price.  Our candidate \emph{coupon profiles} are shown in Table~\ref{tbl:coupon}, where coupons are awarded in periods $1$ and $6$ only if the usage is greater than base usage values of $x_1=2.464$ kWh and $x_6 = 2.24$ kWh, respectively.  We experimented with a range of coupon profiles to explore their impact on the MFE, and present some examples C$_0$--C$_3.$} 



\begin{table}[ht]
\vspace{-0.1in}
 \tbl{{Mean Day-ahead Price and Energy Coupon Profiles}}{
\vspace{-0.05in}
 \begin{tabu}{|c||c|c|c|c|c|c|}
\hline
Index &  Period	& Price/MWh& {C$_0$/kWh} & {C$_1$/kWh}    & C$_2$/kWh    & {C$_3$/kWh}\\ \hline
1 & $2-3$ PM & $\$47$ & $107$ &$100$     &$100$    &$100$           \\
2 & $3-4$ PM &$ \$55$	 &  $ 5.4$ &$0$    &$0$         &$0$            \\
3 & $4-5$ PM &$ \$78$  &   $1.8 $   &$0$    &$0$        &$0$            \\
4 & $5-6$ PM & $\$99.6$  &   $0$    &$0$   &$0$         &$0$             \\
5 & $6-7$ PM &$\$66.5$	 &$3.6$   &$0$  &$0$          &$0$          \\
6 & $7-8$ PM &$\$49.5$	 &$54$    &$0$  &$20$           &$40$       \\  \hline
 \end{tabu}}
 \vspace{-0.1in}
 \label{tbl:coupon}
\end{table}

Given the coupon placement by the LSE, the customers need to determine the number of coupons resulting from each action, and use these values to estimate the utility that they would attain.  Our six actions are shown in Table~\ref{tbl:actions} with their attendant costs and number of coupons received.  The LSE conducts an lottery each week across clusters of $M=50$ homes participating in each lottery.  For each cluster, there is $K=1$ prize for winning the lottery.  We assume that the customers choose the same action on each day of the week, and then participate in the lottery.

\subsection{Utility and Surplus}
As described in Section~\ref{sec:mfe-model}, the user state consists of his/her surplus.  Any rewards result in an increase in surplus by the reward amount $w$, whereas performing an action but not receiving a reward results in decreasing the surplus by some amount $l.$  Since a reward is assured for each action in the benchmark incentive scheme, there are no surplus decrease events.  However, in the lottery scheme, a user that does not win the lottery would see a decrease in surplus.  We select $l$ that results from losing at a lottery to be the average reward obtained form the lottery assuming that every player has an equal probability of winning.  

For the customer utility, which maps surplus to utility units, we use the value function of prospect model (defined in~(\ref{eqn:utility-fn})), $u(x)=x^{\gamma}$ if $x\geq 0$ or $u(x)=-{\varphi} (-x)^{\gamma}$ if $x<0$, where $\varphi=2.25$ and $\gamma=0.88$ according to the empirical studies conducted in \cite{tversky1992advances,kahneman1984choices}. The utility model applies to both the benchmark and the lottery scheme.  Under this model, we expect a user who has lost a number of lotteries to stop participating in the system, since his surplus becomes negative and he is not receiving enough of an incentive to stay, given the cost he bears each day.  Similarly, a user who has won too many times would have a large surplus, and would also not be keen on participating since the marginal utility he gets may not be high enough for him.  The latter observation applies to the benchmark as well, although given the small rewards, we do not expect it to happen frequently.

The participants in the lottery scheme see a distorted probability of winning, parameterized by $\xi,$ as defined in~(\ref{eq:weighting}).  This is an important feature of our model, since it captures the attractiveness of lotteries in incentivizing risky actions.  Consistent with empirical studies in \cite{prelec1998probability}, we choose $\xi=0.37.$

We assume that a customer remains in the system with probability $0.92,$ \emph{i.e.,} the average lifetime is $12$ time steps, which parallels the fact that the main summer season lasts for about three months. Further, a newly entering customer has zero surplus.

\subsection{Equilibria Attained by Incentive Schemes}\label{sec:lotteryscheme}

\subsubsection{Benchmark Scheme}
As described in Section~\ref{sec:benchmark}, we construct a fixed-reward type of incentive scheme to obtain a benchmark with which to compare the performance of the lottery scheme.  Under the benchmark scheme, customers are awarded some percentage of the expected savings that their action is likely to yield to the LSE, shown in Table~\ref{tbl:actions}.  The actual action chosen by the customer will be determined using a dynamic program (DP) similar to the one defined in (\ref{optimalequ}).  
However, since rewards are deterministic, there is no dependence on the belief over competitors' actions, and the only randomness is from the lifetime of the user.  Thus, solving the DP is straightforward, and we can easily obtain a map between surplus and action for a given reward.  

\subsubsection{Lottery Scheme}\label{sec:numerical-mfe}
We next consider the lottery with $M=50$ competitors.  The win probability $p_{\rho, a}$ is the probability that the coupons generated by action $a$ are greater than those generated by $M-1$ independent actions drawn from $\rho.$  We offer a single prize with value $\$15,$ which implies that the customer expects to win {$\cent 30$} on average by participating, i.e., the decrease in surplus due to losing at the lottery is $l=0.3,$ while the increase in surplus due to winning is $w=15-0.3=14.7.$   


{We start with a uniform action distribution $\rho_0$ as the initial condition.  In each iteration $i$, given the belief $\rho_i$ (action distribution of other players), we first determine the value of each state using the Bellman equation for value (\ref{operator}), with convergence in roughly $50$ steps.  We next determine the stationary surplus distribution, and then map it to the resultant stationary action distribution $\rho_{i+1},$ uniformly choosing all equal value actions. Note that this map $\tilde{\Pi},$ from belief $\rho_i$ to the resultant distribution $\rho_{i+1}$, is a sequential version of the map $\Pi^*$ from Section~\ref{sec:mfe-model}, and $\tilde{\Pi}$ and $\Pi^*$ have the same fixed points.   As described in Section~\ref{sec:mfe-discussion}, the iterative procedure is referred to as the Mckean-Vlasov dynamics.   As discussed in Section~\ref{sec:mfe-discussion}, we also identify the CTMC version of our system, and simulate it using the same procedure above.  Finally, as specified in  Section~\ref{sec:mfe-discussion}, an  important sufficiency condition for convergence is the Lipschitzness of the  map $\tilde{\Pi}.$  We calculate a numerical derivative (for both the DTMC and CTMC versions)}
\begin{equation}
\frac{||\tilde{\Pi}_{i+1}(\rho_{i+1}) - \tilde{\Pi}_i(\rho_i)||_{\text{sup}}}{||\rho_{i+1} - \rho_i||_{\text{sup}}}.
\end{equation}
{ Figure~\ref{fig:onoffaction-energydistri} (\textit{Left}) plots the derivative along a simulated trajectory for both DTMC and CTMC. That they are bounded, indicates the Lipschitz property of the maps.} 

{We found that typically convergence occurs rapidly and reaches within $0.1 \%$ of the final value within $20$ iterations.  The eventual values to which each surplus value converge is the mean field surplus distribution, in which it turns out that customers win at a lottery at most once over an average lifetime of $12$ time intervals, as is to be expected with a cluster size of $50$ customers at each lottery.  The mean field action distribution under the lottery scheme with a $\$15$ reward and the savings attained are shown in Table~\ref{tbl:mfes}. The MFE shifts based on the coupon profile,  but the saving is quite robust to profiles that award comparable numbers of coupons in periods $1$ and $6.$}   

{We observed multiple thresholds at which two actions have identical value.  For example, under coupon profile C$_0,$ there are three threshold surplus values $-19.2$, $1.7$ and $190.3$ at which we have equal probabilities of choosing between actions $0$ and $2,$ between $2$ and $4,$ and between $4$ and $5,$ respectively.}
\begin{table}[ht]
\vspace{-0.1in}
 \tbl{Mean Field Equilibria under \$15 reward (lottery prize)}{
\vspace{-0.05in}
 \begin{tabu}{|c||c|c|c|c|}
 \hline
 Coupon Profile & MFE & Expected Surplus & Expected Value & LSE Saving\\ \hline
{C$_0$} & {$[0.001, 0, 0.81, 0, 0.19, 0]$}     &    $0.3563$   &  $\$ 189.8$ &{$\$77$}\\
{C$_1$} & {$[0.001, 0, 0, 0.584, 0, 0.416]$}  &  $0.3704$    & $\$  203.5$  &{$\$69$}  \\
    C$_2$ & $[0.001, 0, 0.875, 0.124, 0, 0]$   &  $0.3565$  & $\$ 193.1 $ &$\$79$\\
 {C$_3$} & {$[0.001, 0, 0.999, 0, 0, 0]$}   &  $0.3563$ &   $\$ 189.4$   &{$\$79.4$}\\
 \hline
  \end{tabu}}
 \vspace{-0.2in}
 \label{tbl:mfes}
\end{table}



%
%

\subsubsection*{Example}
\begin{figure*}[ht]
\centering
\begin{minipage}{.3\textwidth}
\centering
\includegraphics[width=1\columnwidth]{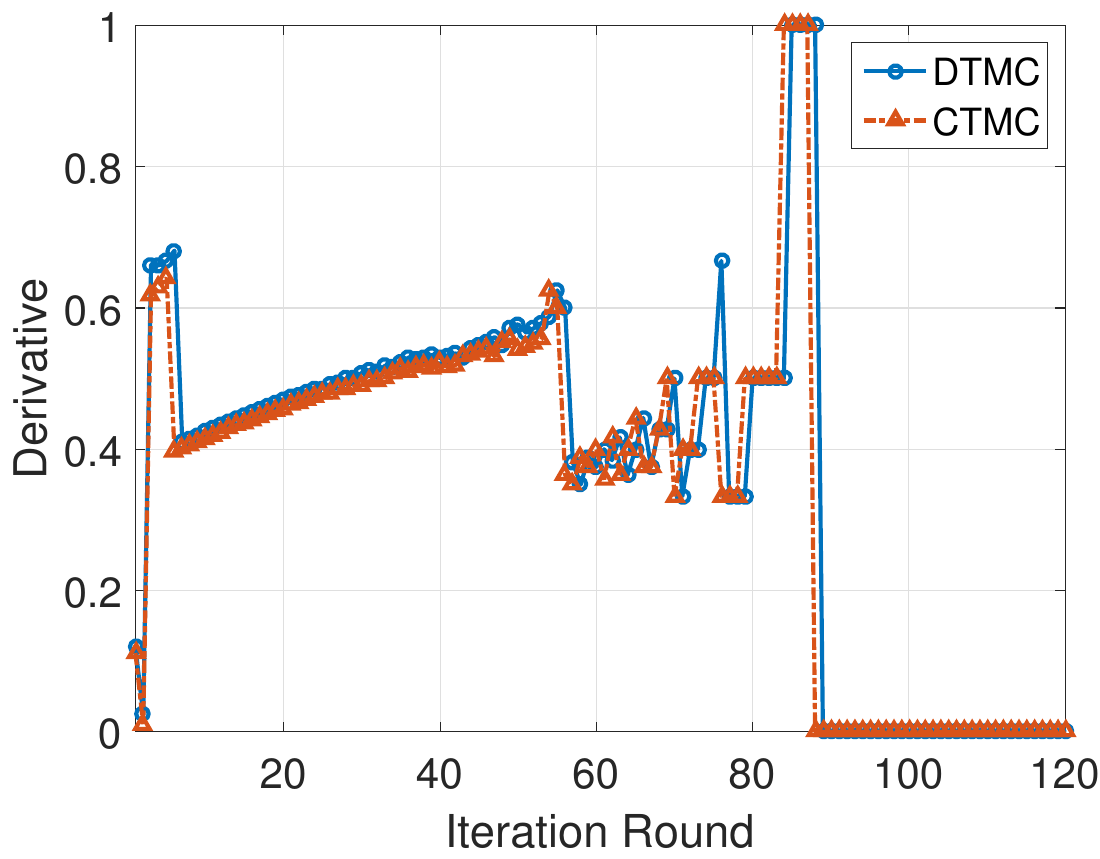}
\end{minipage}\hfill
\begin{minipage}{.3\textwidth}
\centering
\includegraphics[width=1\columnwidth]{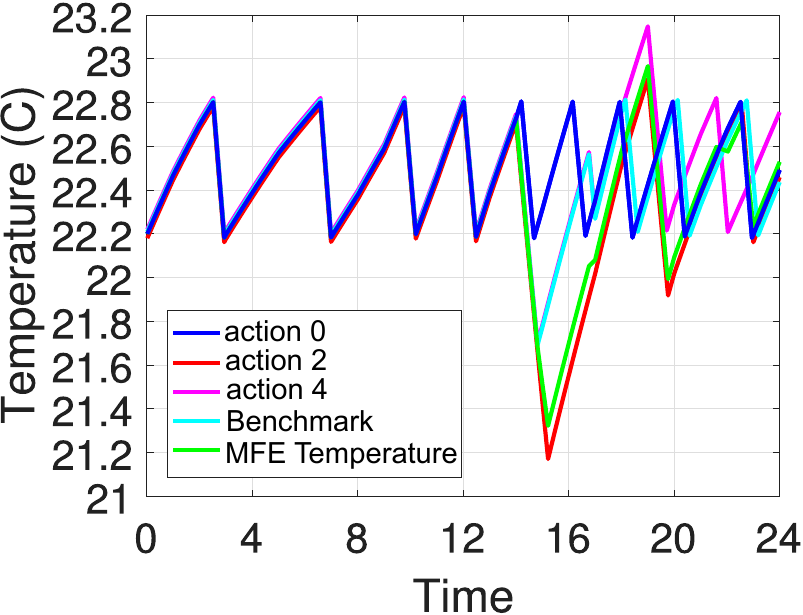}
\end{minipage}\hfill
\begin{minipage}{.3\textwidth}
\centering
\includegraphics[width=1\columnwidth]{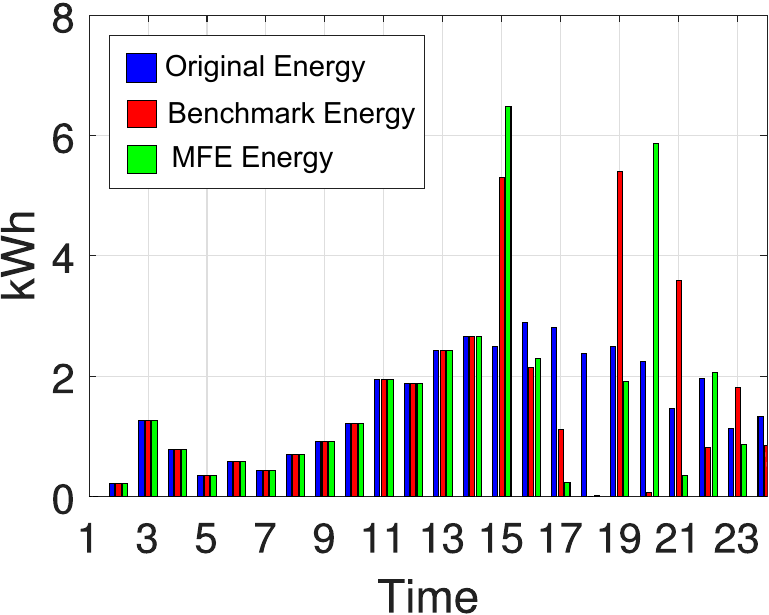}
\end{minipage}
\caption{{{\textit{(Left)} Numerical derivative of map;} \textit{(Middle)} Simulated ON/OFF state of AC over a $24$ hour period under actions $0$, $2$, $4$, the mean field action and the benchmark action on an arbitrary day and the corresponding interior temperature. The temperature graph is slightly offset for actions $2, 4$, the mean field action and the benchmark action for ease of visualization; \textit{(Right)} Average daily energy usage profile.}
\vspace{-0.1in}}
\label{fig:onoffaction-energydistri}
\end{figure*}

Figure~\ref{fig:onoffaction-energydistri} (\textit{Middle}) shows the interior temperature under actions $0$, $2$, $4$, the mean field action distribution and benchmark action when $\$15$ is the total reward amount.   We see that the mean field behavior is more aggressive than the benchmark in reducing the interior temperature before the peak period, and shows a marginally higher interior temperature during the peak period.  Figure~\ref{fig:onoffaction-energydistri} (\textit{Right}) shows the comparison of energy consumption between action $0$ (doing nothing, with an average energy consumption of $36.5$ kWh per day), the mean field action distribution (average energy consumption of $36.7$ kWh per day), and the benchmark action (average energy consumption of $36.4$ kWh per day).  We see that the mean field distribution is more aggressive in moving energy usage away from the peak period as compared to the benchmark, although both have essentially the same energy consumption and an identical reward value of $\$15$  per week.  Finally, we compute that the savings to the LSE over $50$ homes each week in this is example is $\$57.4$ in the benchmark scheme and $\$ 77$ in the lottery scheme.  Thus, incentivizing customers by offering a prize of $\$ 15$ each week is certainly feasible.  The MFE illustrates that even as small as $1\degree$C change of the setpoint of AC each day over several homes can yield significant benefits.




\subsection{Performance Analysis of Incentive Schemes}

\subsubsection*{Benchmark Scheme}
\begin{figure*}[ht]
\centering
\begin{minipage}{.3\textwidth}
\centering
\includegraphics[width=1\columnwidth]{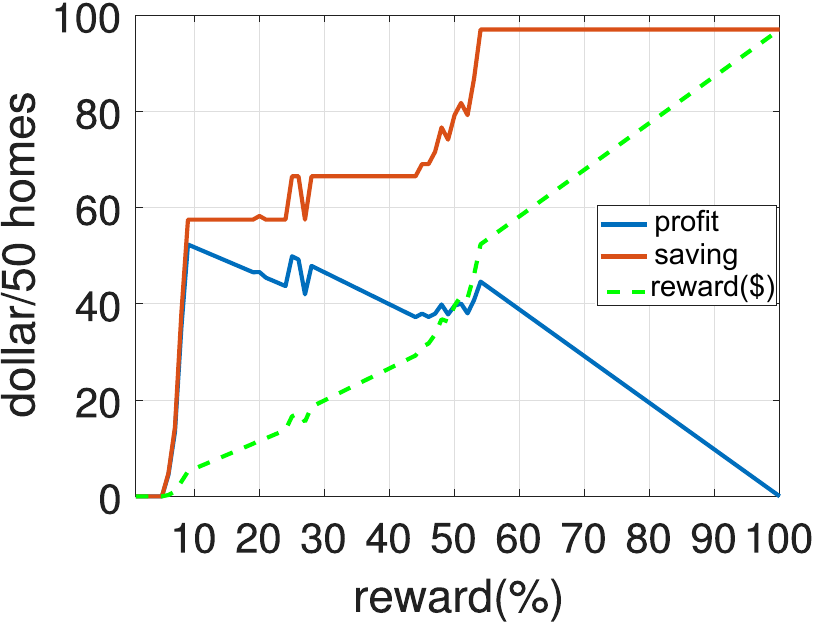}
\end{minipage}\hfill
\begin{minipage}{.3\textwidth}
\centering
\includegraphics[width=1\columnwidth]{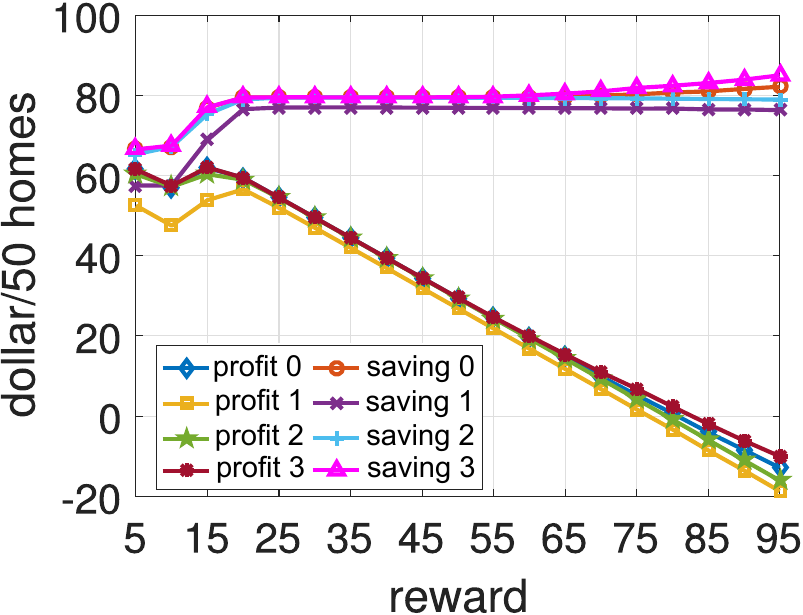}
\end{minipage}\hfill
\begin{minipage}{.3\textwidth}
\centering
\includegraphics[width=1\columnwidth]{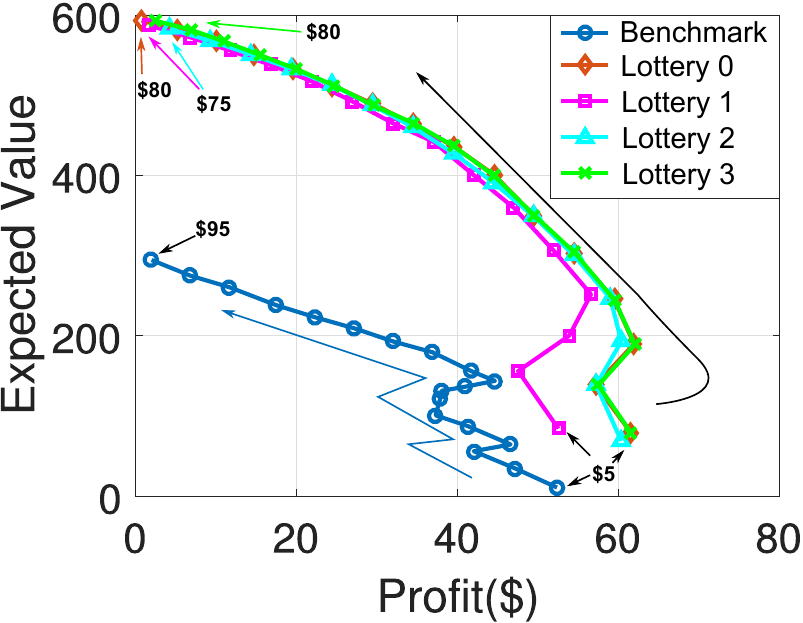}
\end{minipage}
\caption{{The relation between offered reward, LSE savings and LSE profit: \textit{(Left)} Benchmark incentive scheme and \textit{(Middle)} Lottery scheme; \textit{(Right)} The relation between profit to the LSE and the expected value of a generic customer when different rewards are given to the customers.}}
\label{fig:reward-saving-profit-expectedvalue}
\vspace{-0.2in}
\end{figure*}

{We consider a range of scenarios wherein the LSE rewards customers for each action with between $1\%-100\%$ of its expected savings, in steps of $1\%$  increments.  The relations between the total weekly reward to customers, savings to the LSE and profit to the LSE, are shown in {Figure~\ref{fig:reward-saving-profit-expectedvalue} (\textit{Left})}.  We see that the maximum weekly profit of $\$52$ is achieved when about $9\%$ savings (about $\$10$ in total per week) is the customer reward regardless of the coupon awarding profile, indicating robustness to the exact profile employed.   Note that although the reward under the benchmark scheme is indicated by a percentage returned, it corresponds to a dollar value returned based on the actions of the customers, and the total dollar reward values are also shown in green (dashed line) in {Figure~\ref{fig:reward-saving-profit-expectedvalue} (\textit{Left})}. }

\subsubsection*{Lottery Scheme}
{We next conduct numerical experiments under a range coupon profiles and lottery rewards from {$\$5$ to $\$95$} in steps of $\$5$ increments as we did (by using percentage returns) with the benchmark scheme.  Hence, we set a reward value, calculate the values of $l$ and $w$ that it implies, and compute the MFE for different coupon awarding profiles.  Our results are  shown in {Figure~\ref{fig:reward-saving-profit-expectedvalue} (\textit{Middle})}, where we plot the total savings to the LSE as well as its profit (savings minus reward) as a function of the reward offered for winning the lottery.   From {Figure~\ref{fig:reward-saving-profit-expectedvalue} (\textit{Middle})},  the maximum profit is achieved by in a robust manner giving a reward of $\$15-\$20$ for all coupon profiles  {Also, from observation of the mean field action distribution that results from this reward (Table~\ref{tbl:mfes}), we note that almost all the customers will participate in the system, i.e., the probability of choosing action $0$ is close $0$. } }

{From {Figures~\ref{fig:reward-saving-profit-expectedvalue} (\textit{Left}) and~\ref{fig:reward-saving-profit-expectedvalue} (\textit{Middle})}, we see that the maximum profit using the lottery scheme is a little over $\$62$ per week, while the maximum profit is only about $\$52$ under the benchmark scheme.  Given that a typical LSE has several hundred thousand customers, a difference of $\$10$ each week over a cluster of $50$ homes is quite significant.}

\subsubsection*{Comparison of Local Social Welfare}
{Our final step is to  characterize the local social welfare under the lottery (under different coupon profiles) and the benchmark.   We
define the (expected) \emph{local social welfare} (LSW) measured in dollars/week as}
\begin{equation}\label{eq:lsw}
\text{LSW} =50 \mathbb{E}_{\text{MFE}}[V_i(X)] + \text{profits of the LSE}. 
\end{equation}

For the lottery scheme, {the maximum expected value to each user of roughly $600$ each is obtained when the LSE is revenue neutral. This is roughly $100\%$ better than what can be achieved with the benchmark individual incentive.  This increase is due to both the prospect-based utility as well as coupling across users in the lottery scheme.} 

{{Note also that} the LSE obtains a maximum profit between $\$62-\$64.4$ (for different coupon profiles) by giving a $\$15$ reward, under which each customer will take actions according to the MFEs shown in Table~\ref{tbl:mfes}.  The corresponding surplus is also shown, which for a  generic customer is between $\$189.2-\$203.5.$  Therefore, the local social welfare is about $\$9552$.}

For the benchmark, the profit to the LSE is $\$42$ when $\$15$ reward is given as shown in {Figure~\ref{fig:reward-saving-profit-expectedvalue} (\textit{Left}), under which each customer will only take action $5$. } The corresponding expected value of a generic customer is about $\$56.2,$ and the local social welfare is about $\$2852.$  Again, we see that the lottery scheme outperforms the benchmark.

{We perform the same analysis to determine the relation between profit to the LSE and the expected value of a generic customer under different rewards and coupon profiles.   Our results are shown in {Figure~\ref{fig:reward-saving-profit-expectedvalue} (\textit{Right})}.  We explore a range of rewards from {$\$5$ to the break-even point ($\$80$ for lottery scheme and $\$95$ for the benchmark, regardless of the coupons awarded). }  The points on each curve correspond to increasing the reward by {$\$5$} in steps in a manner indicated by the arrow marks.  From {Figure~\ref{fig:reward-saving-profit-expectedvalue} (\textit{Right})}, we see that the lottery scheme appears to better capture the frontier between LSE profit and customer value than the benchmark scheme, {with the lottery-based incentive being better in a Pareto-sense}.  This is true regardless of the exact maximum coupon choice.  Thus, based on a desired level of customer value and LSE profit, the lottery scheme can ensure a better outcome than the benchmark scheme with an appropriate reward.}

\section{Lottery Scheme}\label{sec:lottery}

We first construct the lottery scheme that will be used in our mean field game.  We permute all the agents into clusters, such that there are exactly $M$ agents in each cluster, and conduct a lottery in each such cluster.  Suppose there are $K$ rewards for all agents in one cluster, where $K$ is a fixed number less than $M$.  When an agent takes an action, he/she will receive the credit (number of coupons) associated with that action. Then the probability of winning is based on the number of coupons that each agent possesses. We will model the lotteries as choosing a permutation of the $M$ agents participating in it, and picking the first $K$ of them as winners. Then different lottery schemes can be interpreted as choosing different distributions on the symmetric group of permutations on $M$. In particular, we will use ideas from the Plackett-Luce model to implement our lotteries.  

Without loss of generality, we assume that the actions are ordered in decreasing order of the costs so that $\theta_1 \geq \dotsb \geq \theta_{A}$. In order to incentivize agents to take the more costly actions we will insist that the vector of coupons obtained for each action is also in decreasing order of the index, i.e., $r_1 \geq \dotsb \geq r_{A}$.   

The specific lottery procedure we consider is the following: for every agent $m$ that takes action $a[m]$ and receives coupons $r_{a[m]}>0$, we choose an exponential random variable with mean $1/r_{a[m]}$ and then pick the first $K$ agents in increasing order of the realizations of the exponentials. Note the abuse of notation only in this section to use $a[m]$ to refer to the action of agent $m.$ Since we consider only one lottery, we do not consider time $k$.  Let the agent $m=1,\dotsc, M$ receive $r_{a[m]}$ number of coupons.  The set of winners is a permutation over the agent indices, and we denote such a  permutation by $\mu = [\mu_1, \mu_2,\cdots, \mu_M].$   We then have the probability of the permutation $\mu$ given by
\begin{align}
\mathbb{P}(\mu | r_{a[1]}, \dotsc, r_{a[M]}) = \prod_{n=1}^{M-1} \frac{r_{a[\mu_n]}}{\sum_{j=n}^M r_{a[\mu_j]}}.
\end{align}
Essentially, after each agent is chosen as a winner, he is removed and the next lottery is conducted just as before but with fewer agents. 

We now analyze the probability of winning in our lottery.   For analysis under the mean field assumption, it suffices to consider agent $1$ with the coupons it gets by taking action $a$ being denoted as $r_{a[1]}$. Let $\mathcal{M}:=\{2,\dotsc,M\},$ which is the set of opponents of agent $1.$   For these agents, suppose there are $\upsilon_n$ agents that choose action $n$, where $\sum_{n\in\A}{\upsilon_n}=M-1$.  We denote the vector of these actions by $\vec{\upsilon}=(\upsilon_1, \dotsc, \upsilon_{A}).$   

The conditional probability of agent $1$ failing to obtain a reward is given by
\begin{align*}
p^L_{1,\vec{\upsilon}}=\sum_{\kappa_1\in \mathcal{M}_1} 
\dotsb \sum_{\ka_K \in \mathcal{M}_K} \frac{\prod_{l=1}^K r_{a[\ka_l]}}{\prod_{l=1}^{K} (r_{a[1]} + \sum_{m\in \mathcal{M}_l} r_{a[m]})},
\end{align*}
where $L$ refers to the fact that agent $1$ ``loses,'' $\mathcal{M}_1 = \mathcal{M},$ and for $l \geq 2$ we have $\mathcal{M}_l = \mathcal{M}_{l-1}\setminus \{\ka_{l-1}\}.$  Essentially, the above looks at the lottery process round by round, and is a summation of the probabilities of all permutations in which agent $1$ does not appear in the first spot in any round.

The above expression considerably simplifies if the summations are instead taken over the actions $\tilde{\ka}_l$ that the lottery winner $\ka_l$ at round $l\in \{1, \dotsc, K\}$ can take. Note that we assume that we can distinguish the actions based on the number of coupons given out. If this were not true, then we could further simplify the expression by summing over the coupon space.  Given a coupon/action profile $\vec{\upsilon}$, let $\mathcal{J}(\vec{\upsilon})$ denote the actions that have non-zero entries. Additionally, by $\vec{\upsilon}-\vec{1}_{\tilde{\ka}}$ for $\tilde{\ka}\in\mathcal{J}(\vec{\upsilon})$ denote the resulting coupon profile obtained by removing one entry at location $\tilde{\ka}$, and by $r_{\vec{\upsilon}}$ the sum of all the coupons in profile $\vec{\upsilon}$, i.e., $\sum_{\tilde{\ka}\in \mathcal{J}(\vec{\upsilon})} r_{\tilde{\ka}} \upsilon_{\tilde{\ka}}$. Then 
\begin{align}\label{eq:lossprob}
p^L_{1,\vec{\upsilon}}=\sum_{\tilde{\ka}_1\in \mathcal{J}(\vec{\upsilon}^1)} \dotsb \sum_{\tilde{\ka}_K \in \mathcal{J}(\vec{\upsilon}^K)} \frac{\prod_{l=1}^K \upsilon^l_{\tilde{\ka}_l} r_{\tilde{\ka}_l}}{\prod_{l=1}^K(r_{a[1]}+r_{\vec{\upsilon}^l})},
\end{align}
where $\vec{\upsilon}^1=\vec{\upsilon}$, for $l=2,\dotsc,K$, $\vec{\upsilon}^l = \vec{\upsilon}^{l-1}-\vec{1}_{\tilde{\ka}_l}$ and $\upsilon^l_{\tilde{\ka}}$ is the number of entries at location $\tilde{\ka}$ for coupon profile $\vec{\upsilon}^{l}$. Note that $p^L_{1,\vec{\upsilon}}$ is a decreasing function of $r_{a[1]}$ for every $\vec{\upsilon}$. Therefore, agent $1$ comparing two actions $i$ and $j$ that have $r_{1,i} > r_{1,j}$ will find $p^L_{1,\vec{\upsilon}} (i)< p^L_{1,\vec{\upsilon}}(j)$ for all $\vec{\upsilon}$.  Also by taking the limit of $r_{a[1]}$ going to $0$, having an action with $0$ coupons results in a loss probability of $1$ for every $\vec{\upsilon}$.

To determine the probability of winning in the lottery we need to account for the fact that the actions of the opponents are drawn from the distribution $\rho$ (under the mean field assumption).  Hence, the probability of obtaining the coupon profile (equivalently action profile) of the opponents  $\vec{\upsilon}=(\upsilon_1, \dotsc, \upsilon_{A})$ is given by the multinomial formula,\emph{ i.e.,} 
\BEQA
\label{mfe:realization}
\mathbb{P}_\rho(\vec{\upsilon})=\frac{(M-1)!\prod_{i\in\A} b_i^{\upsilon_i}}{\prod_{i\in\A} \upsilon_i!}.
\EEQA 

Using (\ref{eq:lossprob}) and (\ref{mfe:realization}), we obtain the winning probability for the mean field agent $1$ when taking action $a$ as 
\begin{equation}\label{equ:successprob}
p_{\rho,a} = 1-\sum_{\vec{\upsilon}: |\mathcal{J}(\vec{\upsilon})| = M-1} p^L_{1,\vec{\upsilon}} \mathbb{P}_\rho(\vec{\upsilon}).
\end{equation}
By lower bounding each term in the conditional probability of not obtaining a reward we get $p_{\rho,a} \leq 1- \tfrac{M-K}{M} (\tfrac{r_{A}}{r_1})^K=:\overline{p}_W\in(0,1)$. If we ran the lottery without removing the winners (and any of their coupons), we obtain a lower bound on the probability of winning that has a simpler expression. Using this simpler expression we can obtain the lower bound $p_{\rho,1} \geq 1-(1-\tfrac{r_{A}}{r_A+(M-1)r_1})^K=:\underline{p}_W\in(0,1)$. Note that both bounds are independent of $\rho$. If we allow an action that yields $0$ coupons, then the above bounds become trivial with $\overline{p}_W=1$ and $\underline{p}_W=0$.

An important feature of our lottery scheme is that the probability of winning increases with the number of coupons given out. For simplicity we assumed a fixed reward for any win. However, we can extend the lotteries to ones where different rewards are given out at different stages, and also where the rewards are dependent on the number of coupons of the winner. For the latter, we will insist on the rewards being an increasing function of the number of coupons of the winner. Finally, we can also extend to scenarios where we choose the number of stages $K$ in an (exogenous) random fashion in $\{1, \dotsc, M-1\}$. Since the analysis carries through unchanged except with more onerous notation, we only discuss the simplest setting.

\section{Optimal Value Function}\label{sec:opt-val-fn}
As discussed in Section~\ref{sec:mfe-model}, the mean field agent must determine the optimal action to take, given his surplus $x$ and the assumed action distribution $\rho.$  We follow the usual quasi-linear combination of prospect function and cost consistent with Von Neumann-Morgenstern utility functions, and under which the impact of winning or losing in the lottery is on the surplus of the agent (and not simply a one-step myopic value change). 

The objective of a particular agent $i$ is
\begin{align*}
V_{\rho}(\boldsymbol x[k])=\max_{\left\{\boldsymbol a(\boldsymbol x[l])\in\A^{|\A|}\right\}_{l=k}^{\infty}}\mathbb{E}\left\{\sum_{l=k}^{\infty}\beta^{l-k} \left(u_i(x_i[l])-\theta_{a_i(x_i[l])}\right)\right\},
\end{align*}
where $\boldsymbol x[k]=(x_1[k], \cdots, x_M[k]),$ and $\boldsymbol a(\boldsymbol x[k])=(a_1(x_1[k]), \cdots, a_M(x_M[k]))$ are the vectors of surplus and actions for each agent in the particular lottery cluster of the agent at time $k,$ respectively.  The expectation is over the distribution of competitors actions and the randomness introduced by the lottery.  

Under the mean field assumption, the actions of all agents besides $i$ are drawn from a distribution $\rho$ independently of each other.  Also, the agent uses a prospect function to estimate the probabilities of winning and losing at the lottery.   We can then drop the index of the agent $i$ and the dynamic program that the agent in prospect theory needs to solve is given by the following Bellman equation
\begin{equation}
\begin{aligned}
V_{\rho}(x)=\max_{a(x)\in \A} \{u(x)-\theta_{a(x)}+\beta [\phi(p_{\rho, a}(x))V_\rho(x+w)
+\phi(1-p_{\rho, a}(x))V_\rho(x-l)]  \}.
\label{optimalequ}
\end{aligned}
\end{equation}
Note that $p_{\rho, a}(x)$ is a result of a lottery that we described in detail in Section \ref{sec:lottery}, and $\phi(\cdot)$ is the weighting function, which overweights small probabilities (of winning the lottery) and underweights moderate and high probabilities (of losing the lottery). Here, we use the weighting function defined in~(\ref{eq:weighting}). 

First, we need to define a set of functions as 
\begin{equation*}
\Phi=\left\{f : \X\rightarrow \R: \sup_{x \in \X } \left|\frac{f(x)}{\Omega(x)}\right|<\infty\right\},
\end{equation*}
where $\Omega(x)=\max\{|u(x)|, 1 \}$. Note that $\Phi$ is a Banach space with $\Omega-$norm, 
\begin{equation*}
||f||_{\Omega}=\sup_{x\in \X} \left|\frac{f(x)}{\Omega(x)}\right|<\infty.
\end{equation*}

Also define the Bellman operator $T_\rho$ as
\begin{equation} 
T_\rho f(x)=\max_{a(x)\in \A}\{u(x)-\theta_{a(x)}+\beta [\phi(p_{\rho, a}(x))f(x+w)
+\phi(1-p_{\rho, a}(x))f(x-l)]\},
\label{operator}
\end{equation}
where $f\in\Phi$.

We now show that the optimal value function $V_{\rho}(x)$ exists and it is continuous in $\rho$.

\begin{lemma}\label{lem:Tfp}
1) There exists a unique $f^*\in \Phi$, such that $T_\rho f^*(x)=f^*(x)$ for every $x\in\X$, and given $x\in\X$, for every $f\in \Phi$, we have $T^n_\rho f(x)\rightarrow f^*(x)$, as $n\rightarrow \infty$.\\
2) The fixed point $f^*$ of operator $T_\rho$ is the unique solution of Equation~\eqref{optimalequ}, i.e. $f^*=V_{\rho}^*$.
\end{lemma}

\begin{lemma}\label{lem:Vrhoc}
The value function $V_\rho(\cdot)$ is Lipschitz continuous in $\rho$.
\end{lemma}

\subsection{Stationary distributions}
For a generic agent, w.l.o.g., say agent 1, we consider the state process $\{x_1[k]\}_{k=0}^{\infty}$. It's a Markov chain with countable state-space $\X$, and it has an invariant transition kernel given by a combination of the randomized policy $\sigma(x)$ at each surplus $x$ for any $a(x)\in\A$, and the lottery scheme from Section~\ref{sec:lottery}. By following this Markov policy, we get a process $\{W[k]\}_{k=0}^\infty$ that takes values in $\{\text{win},\text{lose}\}$ with probability $p_{\rho, a}(x)$ for the win, drawn conditionally independent of the past (given $x_1[k]$). Then the transition kernel conditioned on $W[k]$ is given by
\begin{equation}
\pr(x_1[k]\in B|x_1[k-1]=x, W[k])
=\beta 1_{\{x+w 1_{\{W[k]=\text{win}\}}-l 1_{\{W[k]=\text{lose}\}}\in B\}}+(1-\beta)\Psi(B),
\label{eq:transtion}
\end{equation}
where $B\subset \X$ and $\Psi$ is the probability measure of the regeneration process for surplus. The unconditioned transition kernel is then
\begin{align}\label{eq:uncondkernel}
 \pr(x_1[k]\in B|x_1[k-1]=x)  = 
 &\beta \sum_{a(x)\in\A} \sigma_a(x) p_{\rho, a}(x) 1_{x+w\in B} \\\nonumber
 &+ \beta \big(1-\sum_{a(x)\in\A} \sigma_a(x) p_{\rho, a}(x)\big) 1_{x-l\in B} + (1-\beta) \Psi(B).
\end{align}

\begin{lemma}\label{eq:ksteptransitionkernel}
The Markov chain where the action policy is determined by $\sigma(x)$ based on the states of the users and the transition probabilities in~(\ref{eq:uncondkernel}) is positive recurrent and has a unique stationary surplus distribution. We denote the unique stationary surplus distribution as $\zeta_{\rho\times\sigma}$. Let $\zeta_{\rho\times\sigma}^{(k)}(B|x)$ be the surplus distribution at time $k$ induced by the transition kernel~(\ref{eq:uncondkernel}) conditioned on the event that $X[0]=x$ and there is no regeneration until time $k$. $\zeta_{\rho\times\sigma}(\cdot)$ and $\zeta_{\rho\times\sigma}^{(k)}(\cdot)$ are related as follows:
\begin{equation}\label{eq:transitionrelation}
\begin{aligned}
\zeta_{\rho\times\sigma}(B)&=\sum_{k=0}^{\infty}(1-\beta){\beta}^k \mathbb{E}_{\Psi}\left(  \zeta_{\rho\times\sigma}^{(k)}(B|X)\right)=\sum_{k=0}^{\infty}(1-\beta){\beta}^k\int\zeta_{\rho\times\sigma}^{(k)}(B|x)d\Psi(x).
\end{aligned}
\end{equation}
\end{lemma}
Thus $\zeta_{\rho\times\sigma}(B)$ in terms of  $\zeta_{\rho\times\sigma}^{(k)}(B|x)$ is simply based on the properties of the conditional expectation. And note that in $\mathbb{E}_{\Psi}\left(\zeta_{\rho\times\sigma}^{(k)}(B|X)
\right)$, the random variable $X$ is the initial condition of the surplus, distributed as $\Psi$. For $x\in \mathbb{X}$, the only possible one-step updates are the increase of the surplus to $x+w$ or a decrease to $x-l$, i.e. $B=\{x+w, x-l\}.$

\section{Mean Field Equilibrium}\label{sec:mfe-definition}
The action distribution $\rho$ is a probability mass function on the action set $\A$: let $b_i$ be the probability of choosing action $i$. Note that $\rho$ lives in the probability simplex on $\mathbb{R}^{|\A|}$, which is compact and convex; denote it as $\Gamma_{\rho}$. Let $\zeta$ be the stationary surplus distribution and the set of all such possible surplus distributions is denoted as $\Gamma_{\zeta}$, which is a compact and convex subset of $l_\infty$: all surplus distributions are dominated by the distribution obtained by allowing the agent to win in every period; all surplus distributions dominate the distribution obtained by allowing the agent to lose in every period; both these distributions have a finite mean (and convexity follows); and then the compactness result follows using the argument in \cite{LiBha15arxiv}.  For a given surplus $x$, let $\sigma(x)$ be the action distribution at $x$. Denote $\Gamma_{\sigma}$ as the set of all possible distributions over the action space for each $x$, which is compact and convex. We further assume that $\rho\in\Gamma_{\rho}$, $\zeta\in\Gamma_{\zeta}$ and $\sigma(x)\in\Gamma_{\sigma}$ for each $x\in\X$.

\begin{definition}\label{mfe:def}
Consider the action distribution $\rho$, the randomized policy $\sigma$ and the stationary surplus distribution $\zeta_{\rho}$: (i) Given the action distribution $\rho$, determine the success probabilities in the lottery scheme using~(\ref{equ:successprob}) and then compute the value function in~(\ref{optimalequ}). Taking the best response given by~(\ref{optimalequ}) results in an action distribution $\tilde{\sigma};$ (ii) Given action distribution $\rho$, following the randomized policy $\sigma$ yields transition kernels for the surplus Markov chain and stationary surplus distribution $\tilde{\zeta}_{\rho}$, (with each transition kernel having a unique stationary distribution); and (iii) Given the stationary surplus distribution $\zeta_{\rho}$, applying the randomized policy $\sigma(x)$ at each surplus $x$ yields the distribution of actions $\tilde{\rho}.$ Define the best response mapping $\Pi^*$ that maps $\Gamma_{\rho}\otimes\Gamma_{\sigma}^{|\mathbb{X}|}\otimes\Gamma_{\zeta}$ into itself. Then we say that the assumed action distribution $\rho$, randomized policy $\sigma$ and stationary surplus distribution $\zeta_{\rho}$ constitute a mean field equilibrium (MFE) if $\Pi^*: \rho\otimes\sigma\otimes\zeta_{\rho}\mapsto\tilde{\rho}\otimes\tilde{\sigma}\otimes\tilde{\zeta}_{\rho}$ has $(\rho, \sigma, \zeta_{\rho})$ as a fixed point.
\end{definition}

\subsection{Existence of MFE}\label{sec:mfe-existence}
\begin{theorem}
There exists an MFE of $\rho$, the randomized policy $\sigma(x)$ at each surplus $x$ and $\zeta$, such that $\rho\in\Gamma_{\rho}$, $\sigma(x)\in\Gamma_{\sigma}$ and $\zeta\in\Gamma_{\zeta}$, $\forall a\in\A$ and $\forall x\in\X$.
\end{theorem}

We will be specializing to the spaces $\Gamma_{\rho}, \Gamma_{\sigma}, \Gamma_{\zeta}$ and define the topologies being used in the following proofs first.
\begin{enumerate}
\item  For the assumed action distribution $\rho\in\Gamma_{\rho}$ on the finite set $\mathcal{A}$, all norms are equivalent, we will consider the topology of uniform convergence, i.e., using the $l_{\infty}$ norm given by $||\rho||=\max_{a\in\mathcal{A}}\rho(a)$.
\item For the randomized policy $\sigma\in\Gamma_{\sigma}^{|\mathbb{X}|},$ we enumerate the elements in $\mathbb{X}$ as $1, 2, \cdots,$ and consider the metric topology generated by norm $||\sigma||=\sum_{j=1}^{\infty}2^{-j}|\sigma(x_j)|$, where $|\sigma(x)|=\max_{a\in\mathcal{A}}\sigma(x, a).$ We consider the convergence of any sequence $\{\sigma_n\}_{n=1}^{\infty}$ to $\sigma$ in this topological space.
\item For the surplus distribution $\zeta$ on the countable set $\mathbb{X}$, we consider the topology of pointwise convergence, which can be shown to be equivalent to convergence in $l_\infty$, i.e., uniform convergence, using coupling results presented in \cite{LiBha15arxiv}.
\end{enumerate}

Note that from the definition of $\Gamma_{\rho}$, $\Gamma_{\sigma}$ and $\Gamma_{\zeta}$, they are already non-empty, convex and compact. Furthermore, they are jointly convex. Then in order to show that the mapping $\Pi^*$ satisfies the conditions of Kakutani fixed point theorem, we only need to verify the following three lemmas.

\begin{lemma}\label{claim2}
{ Given $\rho$, by taking the best response given by~(\ref{optimalequ}), we can obtain the action distribution $\sigma(x)$ for every $x$, which is upper semicontinuous in $\rho$.}
 \end{lemma}

\begin{remark}
{As we have discussed earlier, since our state space and action space are  discrete,  there might exist multiple best response actions when the agent solves the dynamic program.  Thus, a pure equilibrium might not exist.  Since the best response can be set-valued, we need to consider mixed strategies.  In other words, the agent needs to choose a randomized action policy for each state.   Hence, the randomized policy $\sigma$ is critical in the construction of the MFE given in Definition~\ref{mfe:def}. }
\end{remark}

\begin{lemma}\label{claim3}
Given $\rho$ and $\sigma(x)$, there exists a unique stationary surplus distribution $\zeta(x)$, which is continuous in $\rho$ and $\sigma(x)$.
\end{lemma}

\begin{lemma}\label{claim1}
Given $\zeta(x)$ and $\sigma(x)$, there exists a stationary action distribution $\rho$, which is continuous in $\zeta(x)$ and $\sigma(x)$.
\end{lemma}

\section{Characteristics of the Best Response Policy}\label{sec:best-resp}

In this section, we characterize the best response policy under the assumption that $V_{\rho}$ in~(\ref{optimalequ}) has some properties. Then we discuss the relations between the incremental utility function $u(x)$ and the optimal value function $V_{\rho}$.

\subsection{Existence of Threshold Policy}\label{thresholdpolicy}
We make the assumption that given the action distribution $\rho$, $V_{\rho}(x)$ is increasing and submodular in $x$ when $x\leq -l$; increasing and linear in $x$ when $-l\leq x\leq w$; and increasing and supermodular in $x$, when $x\geq w$. 

In Section~\ref{sec:lottery}, our lotteries are constructed such that the probability of winning monotonically increases with the cost of the action. This when combined with the monotonicity,  submodularity (decreasing differences) for positive argument and supermodularlity (increasing differences) for negative argument of $V_\rho$ yields the following characterization of the best response policy.
\begin{lemma}\label{lem:thresholdpolicy}
For any two action, say actions $a_1$ and $a_2$, suppose that $\theta_{a_1}>\theta_{a_2}$, so that $p_{\rho, a_1}>p_{\rho, a_2}$, i.e., $\phi(p_{\rho, a_1})>\phi(p_{\rho, a_2})$, then there is a threshold value of the surplus queue for user such that preference order for the actions changes from one side of the threshold to the other.
\end{lemma}

Using the same argument as Lemma~\ref{lem:thresholdpolicy}, under the assumption that $V_{\rho}(x)$ is increasing and submodular in $x\in(-\infty, \infty)$, or increasing and supermodular in $x\in(-\infty, \infty)$,  we can show the existence of a threshold policy.

\subsection{Relations between incremental utility function $u(x)$ and the optimal value function $V_\rho$}

\subsubsection{{Concave/Convex incremental utility function}}
{\begin{lemma}\label{Vcontinuous}
Given the action distribution $\rho$, $V_{\rho}(x)$ is an increasing and submodular (i.e., decreasing differences) function of $x$ if $u(x)$ is a concave and monotone increasing function of $x$, supermodular (i.e., increasing differences) function of $x$ if $u(x)$ is a convex and monotone increasing function of $x$.
\end{lemma}}

{Thus, from Lemma~\ref{Vcontinuous}  and Lemma~\ref{lem:thresholdpolicy}, the optimal policy takes a threshold form for both concave and convex incremental utility function.}

\subsubsection{{Conjecture for S-shaped prospect incremental utility function}}

 {We found numerically that with an S-shaped utility function, the value function satisfies the super/sub-modularity conditions on the positive/negative axis respectively.  If this holds true in general, then from  Lemma~\ref{lem:thresholdpolicy}, the optimal policy would take a threshold form, and indeed this is what we observed numerically.   However, we are not able to prove this result due to the implicit nature of the value function, and we can only conjecture that this condition might hold for some class of  S-shaped utility functions.  
}

\section{Conclusion}\label{sec:conclusion}
In this paper we developed a general framework for analyzing incentive schemes, referred to as nudge systems, to promote desirable behavior in societal networks by posing the problem in the form of a Mean Field Game (MFG). Our incentive scheme took the form of awarding coupons in such that higher cost actions would correspond to more coupons, and conducting a lottery periodically using these coupons as lottery tickets.  Using this framework, we developed results in the characteristics of the optimal policy and showed the existence of the MFE.

We used the candidate setting of an LSE trying to promote demand-response in the form of setting high setpoints in higher price time of the day in order to transfer energy usage from a higher to a lower price time of day for an air conditioning application. We conducted data driven simulations that accurately account for electricity prices, ambient temperature and home air conditioning usage.  We showed how the prospect of winning at a lottery could potentially motivate customers to change their AC usage patterns sufficiently that the LSE can more than recoup the reward cost through a likely reduced expenditure in electricity purchase.  Further, we showed that a lottery is more effective than a fixed reward at enabling such desirable behavior and can attain a better tradeoff between social value and LSE profits.

\appendix
\section*{APPENDIX}

\section{Properties of the optimal value function}
{\bf{Proof of Lemma~\ref{lem:Tfp}}}\label{sec:PutermanLemma}

We first show that $T_\rho f\in \Phi$ for $\forall f\in \Phi$. The proof then follows through a verification of the conditions of Theorem $6.10.4$ in \cite{Put94}. From the  definition of $T_\rho$ in~(\ref{operator}), we have
\begin{equation*}
|T_\rho f(x)| \leq |u(x)| + \max_{a(x)\in\A} \theta_{a(x)} +\beta \max(|f(x+w)|,|f(x-l)|).
\end{equation*}
From this it follows that
\begin{align*}
 \sup_{x\in \X} \frac{|T_\rho f(x)|}{\Omega(x)}  \leq \sup_{x\in \X} \frac{|u(x)|}{\Omega(x)} + \sup_{x\in \X} \frac{\max_{a(x)\in\A} \theta_{a(x)}}{\Omega(x)} + \beta \max\left(\sup_{x\in \X} \frac{|f(x+w)|}{\Omega(x)}, \sup_{x\in \X} \frac{|f(x-l)|}{\Omega(x)}\right).
\end{align*}

Let $x_+$ be the unique positive surplus such that $u(x_+)=1$ and $x_-$ be the unique negative surplus such that $u(x_-)=-1$. Note that $\Omega(x)$ is non-decreasing for $x\geq x_-$ and non-increasing for $x\leq x_+$. To avoid cumbersome algebra we will assume $x_+-w>0$ and $x_-+l>0$. Since $\Omega(x) \geq |u(x)|\geq 0$ and $\Omega(x) \geq 1$, the first two terms are bounded by $1$ and $\max_{a(x)\in\A} \theta_{a(x)}$. For the last term we have
\begin{align*}
& \sup_{x\in \X} \frac{|f(x+w)|}{\Omega(x)}  \leq \| f\|_\Omega \sup_{x\in \X} \frac{\Omega(x+w)}{\Omega(x)}. 
\end{align*}
We have the following
\begin{align*}
\frac{\Omega(x+w)}{\Omega(x)} = 
\begin{cases}
\frac{u(x+w)}{\max(u(x),1)} \leq \frac{u(x+w)}{u(x)}, & \text{if } x\geq x_+, \\
\frac{u(x+w)}{\max(u(x),1)} \leq \frac{u(x+w)}{u(x_+)}, & \text{if } x\in [x_+-w, x_+], \\
1, & \text{if }  x\in [x_-,x_+-w], \\
\frac{1}{|u(x)|}  \leq 1, & \text{if } x\in [x_--w,x_-],\\
\frac{u(x+w)}{u(x)}  \leq 1, & \text{if } x \leq x_--w.
\end{cases}
\end{align*}
For $x\geq x_+$, we know using monotonicity of $u(\cdot)$ 
\begin{align*}
\frac{u(x+w)}{u(x)} & = 1+ \frac{u(x+w)-u(w)}{w} \frac{w}{u(x)}  \leq 1+ \frac{u(x+w)-u(w)}{w} w.
\end{align*}
Additionally, for $x\in [x_+-w, x_+]$ we have
\begin{align*}
\frac{u(x+w)}{u(x_+)} & = 1+ \frac{u(x+w)-u(x_+)}{x+w-x_+} (x+w-x_+) \leq 1+ \frac{u(x+w)-u(x_+)}{x+w-x_+} w.
\end{align*}
For the analysis we assume that $u(\cdot)$ is Lipschitz such that $\sup_{x\in\X} u^\prime(x) <+\infty$. Therefore, by the mean value theorem  
\begin{align*}
&\frac{u(x+w)-u(x)}{w} =u^\prime(\xi_1) \leq \sup_{x\geq x_+} u^\prime(x), &\forall \xi_1, x\in[x_+,\infty),\\
&\frac{u(x+w)-u(x_+)}{x+w-x_+}=u^\prime(\xi_2)\leq\sup_{x\in [x_+-w, x_+]}  u^\prime(x), &\forall \xi_2, x\in[x_+-w, x_+],\\
&\sup_{x\in \X}\frac{\Omega(x+w)}{\Omega(x)} \leq \| f\|_\Omega ( 1+ w \sup_{x\geq x_+-w} u^\prime(x) ), &\forall x\in[x_+-w, \infty).
\end{align*}
Similarly, we have 
\begin{align*}
\sup_{x\in \X} \frac{|f(x-l)|}{\Omega(x)} \leq \| f\|_\Omega \sup_{x\in\X} \frac{\Omega(x-l)}{\Omega(x)}.
\end{align*}
Now we have the following
\begin{align*}
\frac{\Omega(x-l)}{\Omega(x)} = 
\begin{cases}
\frac{u(x-l)}{\min(u(x),-1)} \leq \frac{u(x-l)}{u(x)}, & \text{if } x\leq x_-, \\
\frac{u(x-l)}{\max(u(x),-1)} \leq \frac{u(x-l)}{u(x_-)}, & \text{if } x\in [x_-, x_-+l], \\
1, & \text{if }  x\in [x_-+l,x_+], \\
\frac{1}{u(x)}  \leq 1, & \text{if } x\in [x_+,x_++l],\\
\frac{u(x-l)}{u(x)}  \leq 1, & \text{if } x \leq x_+l.
\end{cases}
\end{align*}
Using the same logic as before, we get
\begin{align*}
\sup_{x\in\X} \frac{\Omega(x-l)}{\Omega(x)} \leq \|f\|_\Omega (1+l \sup_{x\in\X:x \leq x_-} u^\prime(x)).
\end{align*}
Since $u(\cdot)$ is Lipschitz, thus, there exists an $\alpha_0\in (0,+\infty)$ such that $\| T_\rho f \|_\Omega \leq \alpha_0$.

Next, we need to verify the conditions of Theorem 6.10.4 in \cite{Put94}. The lemma requires verification of the following three conditions. We set $x[k]$ to be the state variable denoting the surplus at time $k$. We need to show that $\forall x\in \X$, for some constants (independent of $\rho$) $\alpha_1>0$, $\alpha_2>0$ and $0<\alpha_3<1$, 
\begin{equation}
\sup_{a(x)\in \A} |u(x)-\theta_{a(x)}|\leq \alpha_1 \Omega(x),
\label{eq:cond1}
\end{equation}
\begin{equation}
\E_{x[1],a_0} [\Omega(x[1]) | x[0]=x]\leq \alpha_2 \Omega(x), \quad \forall a_0 \in \A,
\label{eq:cond2}
\end{equation}
with the distribution of $x[1]$ chosen based on action $a_0$, and
\begin{equation}
{\beta}^J \E_{x[J],a_0,a_1,\dotsc,a_{J-1}} [\Omega(x[J])| x[0]=x]\leq \alpha_3 \Omega(x), 
\label{eq:cond3}
\end{equation}
for some $J>0$ and all possible action sequences, i.e., $a_j\in\A$ for all $j=0, 1, \dotsc, J-1$ with the distribution of $x[J]$ chosen based on the action sequence $(a_0,a_1,\dotsc, a_{J-1})$ chosen.

First consider (\ref{eq:cond1}). Since $\Omega(x)=\max(|u(x)|,1)$, using the earlier analysis in Section~\ref{sec:lottery}, (\ref{eq:cond1}) is true with $\alpha_1=1+\max_{a\in\A}\theta_a$. Now consider (\ref{eq:cond2}). We have 
\begin{equation*}
\begin{aligned}
 \E_{x[1],a_0}[\Omega(x[1])|x[0]=x] & = \E_\rho[\phi(p_{\rho,a}(x)) \Omega(x+w)+\phi(1-p_{\rho,a}(x)),\Omega(x-l)] \\
&\leq \max(\Omega(x+w),\Omega(x-l)), 
\end{aligned}
\end{equation*}
which is bounded by $\alpha_2 \Omega(x)$ using our analysis from before.

Finally, ~(\ref{eq:cond3}) holds true using the properties of $\Omega(\cdot)$, the bounds on the probability of winning and losing {(from Section \ref{sec:lottery})} and our analysis from earlier in the proof as follows:
\begin{equation*}
\begin{aligned}
& {\beta}^J \E_{x[J],a_0,a_1,\dotsc,a_{J-1}} [\Omega(x[J])| x[0]=x]\\
\leq &{\beta}^J \max(\phi(\overline{p}_W),\phi(1-\underline{p}_W))^J \max(\Omega(x+J w),\Omega(x-J l)) \\
 \leq& (\beta \max(\phi(\overline{p}_W),\phi(1-\underline{p}_W))^J \alpha_4(J) \Omega(x),
\end{aligned}
\end{equation*}
for some affine $\alpha_4(J)>0$ using our analysis from before. 
It now follows that take $J$ large enough we obtain an $\alpha_3 <1$ that is also independent of $\rho$. Note that we can get a simpler bound of 
\begin{align*}
 {\beta}^J \E_{x[J],a_0,a_1,\dotsc,a_{J-1}} [\Omega(x[J])| x[0]=x] \leq \beta^J \alpha_4(J) \Omega(x),
\end{align*}
using just the properties of $\Omega(\cdot)$. Again we can take $J$
large enough to obtain a $\alpha_3 < 1$ that is independent of
$\rho$. This bound is useful when there is an action for which the
probability of winning or losing is $1$. Since all the conditions of
Theorem 6.10.4 of \cite{Put94} are met, then the first result in the lemma holds true. The second then follows immediately from \eqref{optimalequ}.

{\bf{Proof of Lemma~\ref{lem:Vrhoc}}}\label{sec:Vcontinuous}

For any given $\rho$, from Lemma~\ref{lem:Tfp} we know that there is a unique $V_\rho(\cdot)$. Furthermore, it is the unique fixed point of operator $T_\rho$ where $T_\rho^J$ is a contraction mapping with constant $\alpha_3$ that is independent of $\rho$. From \eqref{operator}, it follows that $T_\rho^J$ is a continuous in $\rho$: computing derivatives using the envelope theorem and the expressions from Section~\ref{sec:lottery}, it is easily established that $T_\rho^J$ is, in fact, Lipschitz with constant $(M-1)^J$ when the uniform norm is used for $\rho$. 

Let $\rho_1$ and $\rho_2$ be two population/action profiles such that $\|\rho_1-\rho_2\| \leq \epsilon$ (the choice of norm is irrelevant as all are equivalent for finite dimensional Euclidean spaces). As $T^J_\rho$ is continuous in $\rho$, there exists a $\delta>0$ such that $\|T^J_{\rho_1}V_{\rho_2} - T^J_{\rho_2} V_{\rho_2}\|_\Omega \leq \delta$. However, since $T^J_{\rho_2}V_{\rho_2}=V_{\rho_2}$, we have shown that $\|T^J_{\rho_1} V_{\rho_2} - V_{\rho_2}\|_\Omega \leq \delta$. Applying $T^J_{\rho_1}$ $n$ times and using the contraction property of $T^J_{\rho_1}$, we get
\begin{align*}
\| T_{\rho_1}^{(n+1)J} V_{\rho_2} - T_{\rho_1}^{n J} V_{\rho_2} \|_\Omega \leq \alpha^n_3 \delta.
\end{align*}
The proof then follows since $\lim_{n\rightarrow\infty} \| T_{\rho_1}^{nJ} V_{\rho_2} - V_{\rho_1}\|_\Omega=0$ so that
\begin{align*}
\| V_{\rho_1} - V_{\rho_2} \|_\Omega \leq \sum_{n=0}^\infty \| T_{\rho_1}^{(n+1)J} V_{\rho_2} - T_{\rho_1}^{n J} V_{\rho_2} \|_\Omega \leq \frac{\delta}{1-\alpha_3}.
\end{align*}
Furthermore, using the comment from above we can show that $V_\rho$ is Lipschitz continuous in $\rho$.

\section{The existence and uniqueness of stationary surplus distribution}
{\bf{Proof of Lemma~\ref{eq:ksteptransitionkernel}}}\label{sec:Transition-Kernel}

First, from the transition kernel~(\ref{eq:uncondkernel}), we satisfy the Doeblin condition as
\begin{equation*}
\mathbb{P}(x[k]\in B|x[k-1]=x)\geq(1-\beta)\Psi(B),
\end{equation*}
where $0<\beta<1$, and $\Psi$ is a probability measure for the regeneration process.  Then from results in \cite[Chapter 12]{MeyTwe09}, we have a unique stationary surplus distribution. 

Next,  let $-\tau$ be the last time before $0$ that the surplus has a regeneration. Then we have 
\begin{equation}
\begin{aligned}
\zeta_{\rho\times\sigma}(B)=\sum_{k=0}^{\infty}\mathbb{P}(B, \tau=k)
                                           =\sum_{k=0}^{\infty}\mathbb{P}(B|\tau=k)\cdot\mathbb{P}(\tau=k).
\end{aligned}
\end{equation}
Since the regeneration process happens independently of the surplus with inter-regeneration times geometrically distributed with parameter $(1-\beta)$, then $\mathbb{P}(\tau=k)=(1-\beta){\beta}^k$.  Also given $\tau=k$, we have $X_{-k}\sim\Psi$. Therefore
\begin{align}
\zeta_{\rho\times\sigma}(B)&=\sum_{k=0}^{\infty}(1-\beta){\beta}^k\mathbb{P}(B|\tau=k)
                                           =\sum_{k=0}^{\infty}(1-\beta){\beta}^k\mathbb{E}\left(\mathbb{E}\left(1_{x[0]\in B}|\tau=k, X_{-k}=X\right)|\tau=k\right)\nonumber\\
                                           &=\sum_{k=0}^{\infty}(1-\beta){\beta}^k\mathbb{E}\left(\zeta_{\rho\times\sigma}^{(k)}(B|X)|\tau=k\right)
                                           =\sum_{k=0}^{\infty}(1-\beta){\beta}^k\mathbb{E}_{\Psi}\left(\zeta_{\rho\times\sigma}^{(k)}(B|X)\right)\nonumber\\
                                          & =\sum_{k=0}^{\infty}(1-\beta){\beta}^k\int\zeta_{\rho\times\sigma}^{(k)}(B|x)d\Psi(x).           
\end{align}

\subsection{Existence of MFE}
{\bf{Proof of Lemma~\ref{claim2}}}\label{sec:Claim2}

Define the increasing and piecewise linear convex function 
\begin{equation}
\begin{aligned}
g_{\rho}(y)=\max_{a\in\mathcal{A}} \phi(p_{\rho, a})y-\theta_a
                  =\max_{\sigma\in\Delta(|\mathcal{A}|)}\sum_{a\in\mathcal{A}}\sigma_a \big( \phi(p_{\rho, a})y-\theta_a\big),
\end{aligned}
\end{equation}
where $\Delta(\mathcal{A})$ is the probability simplex on $A=|\mathcal{A}|$ elements. By the properties of the lottery and the weight function $\phi(\cdot)$, $\phi(p_{\rho, a})$ is continuous in $\rho$ for all $a\in\mathcal{A}$. Using Berge's maximum theorem, we have 
\begin{equation}
\arg\max_{\sigma\in\Delta(|\mathcal{A}|)}\sum_{a\in\mathcal{A}}\sigma_a \left( \phi(p_{\rho, a})y-\theta_a\right)
\end{equation}
is upper semicontinuous in $\rho$.

Now let 
\begin{equation}
\mathcal{A}(y):=\arg\max g(y)=\arg\max_{a\in\mathcal{A}} \phi(p_{\rho, a})y-\theta_a,
\end{equation}
then set-valued function above is exactly $\Delta(|\mathcal{A}(y)|)$.

Hence, the optimal randomized policies at surplus $x$ are a set-valued function $\Delta(|\mathcal{A}(y)|)=\Delta(|\mathcal{A}(V_{\rho}(x+w)-V_{\rho}(x-l))|)$, which is upper semicontinuous due to the Lipschitz continuity of $V_{\rho}(\cdot)$ in $\rho$ and the u.s.c. of $\phi(p_{\rho, a})$ in $\rho$, i.e., for every state $x$, the action distribution $\sigma(x)$ is (pointwise) upper semicontinuous in $\rho$.

{\bf{Proof of Lemma~\ref{claim3}}}\label{sec:Claim3}

The existence and uniqueness of $\zeta(x)$ for a given $\rho$ and $\sigma(x)$, and the relationship between $\zeta(\cdot)$ and $\zeta^{(k)}(\cdot)$ are shown in Lemma~\ref{eq:ksteptransitionkernel}.
Now, we will prove the continuity of $\zeta_{\rho\times\sigma}$ in $\rho$ and $\sigma(x)$ for every surplus $x\in \mathbb{X}$. For the assumed action distribution $\rho$ on the finite set $\mathcal{A}$, we consider the topology of pointwise convergence which is equivalent to the uniform convergence by strong coupling results in \cite{LiBha15arxiv}.   For the randomized action distribution $\sigma$, corresponding to $\sigma(x)$ at each surplus $x\in \mathbb{X}$, we consider the topology with metric $\rho(\sigma^1,\sigma^2)=\sum_{j=1}^{\infty} 2^{-j} \min(\|\sigma^1(x_j)-\sigma^2(x_j)\|,1)$, where $\|\cdot\|$ is any norm for $\mathbb{R}^{|\mathcal{A}|}$.

First, we will show that the surplus distribution $\zeta_{\rho\times\sigma}^{(k)}$ is continuous in $\rho$ and $\sigma$. By Portmanteau theorem, we only need to show that for any sequence $\rho_n\rightarrow\rho$ uniformly, $\sigma_n\rightarrow\sigma$ pointwise,  and any open set $B$, we have $\liminf_{n\rightarrow\infty} \zeta_{\rho_n\times\sigma_n}^{(k)}(B|x)\geq \zeta_{\rho\times\sigma}^{(k)}(B|x)$.

\begin{lemma}\label{claim3-lemma}
$\liminf_{n\rightarrow\infty} \zeta_{\rho_n\times\sigma_n}^{(k)}(B|x)\geq \zeta_{\rho\times\sigma}^{(k)}(B|x).$
\end{lemma}
{\bf{Proof of Lemma~\ref{claim3-lemma}}}\label{sec:claim3-lemma}

The proof proceeds by induction on $k$. For $k=0,$ $\zeta_{\rho_n\times\sigma_n}^{(0)}(B|x)=1_{(x\in B)}$ is a point-mass at $x$ irrespective of $\rho_n\times\sigma_n$, and in fact, for any $n\in\mathbb{N}_+$, we have $\zeta_{\rho_n\times\sigma_n}^{(0)}(B|x)=\zeta_{\rho\times\sigma}^{(0)}(B|x)$. Let $\rho_n\rightarrow\rho$ uniform, and $\sigma_n(x)\rightarrow\sigma(x)$ pointwise for every surplus $x$. We will show that $\zeta_{\rho_n\times\sigma_n}^{(k)}(B|x)$ converges pointwise to $\zeta_{\rho\times\sigma}^{(k)}(B|x)$. 

We will refer to the measure and random variables corresponding to $\rho_n\times\sigma_n$ for the $n^{\text{th}}$ system and those corresponding to $\rho\times\sigma$  as coming from the limiting system. We will prove that $\zeta_{\rho_n\times\sigma_n}^{(k)}(B|x)$ converges to $\zeta_{\rho\times\sigma}^{(k)}(B|x)$ pointwise using the metrics given above.

Suppose that the hypothesis holds true for $k-1$ where $k >1$, i.e., $\zeta_{\rho_n\times\sigma_n}^{(k-1)}(B|x)$ converges pointwise to $\zeta_{\rho\times\sigma}^{(k-1)}(B|x)$. To prove this lemma, we only need to show that the hypothesis holds for $k$. Let $\mathbb{P}_{\rho\times\sigma, x}(\cdot)$ be the one-step transition probability measure of the surplus dynamics conditioned on the initial state of the surplus being $x$, and there is no regeneration. Then we have $\mathbb{P}_{\rho_n\times\sigma_n, x}(x+w)=\sum_{a\in\sigma_n(x)}p_{\rho_n\times\sigma_n, a}$, $\mathbb{P}_{\rho_n\times\sigma_n, x}(x-l)=1-\sum_{a\in\sigma_n(x)}p_{\rho_n\times\sigma_n, a}$ and $\mathbb{P}_{\rho\times\sigma, x}(x+w)=\sum_{a\in\sigma(x)}p_{\rho\times\sigma, a}$, $\mathbb{P}_{\rho\times\sigma, x}(x-l)=1-\sum_{a\in\sigma(x)}p_{\rho\times\sigma, a}$. By the properties of the lottery, $p_{\rho\times\sigma, a}$ is continuous in $\rho\times\sigma$ for all $a\in\mathcal{A}$, thus we have $p_{\rho_n\times\sigma_n, a}$ converges to $p_{\rho\times\sigma, a}$ pointwise, i.e., $\mathbb{P}_{\rho_n\times\sigma_n, x}(\cdot)$ converges to $\mathbb{P}_{\rho\times\sigma, x}(\cdot)$ pointwise. By the Skorokhod representation theorem \cite{billingsley2013}, there exist random variables $X_n$ and $X$ on common probability space and a random integer $N$ such that $X_n\sim\mathbb{P}_{\rho_n\times\sigma_n,x}(\cdot)$ for all $n\in\mathbb{N}$, and $X\sim\mathbb{P}_{\rho\times\sigma, x}(\cdot)$ , and $X_n=X$ for $n\geq N$. 

Then we have,
\begin{align}
\liminf_{n\rightarrow\infty} \zeta_{\rho_n\times\sigma_n}^{(k)}(B|x)&=\liminf_{n\rightarrow\infty} \mathbb{E}\left(\zeta_{\rho_n\times\sigma_n}^{(k-1)}(B|X_n)\right)\geq  \mathbb{E}\left(\liminf_{n\rightarrow\infty}\zeta_{\rho_n\times\sigma_n}^{(k-1)}(B|X_n) \right)\nonumber\displaybreak[0]\\
&\geq \mathbb{E}\left(\zeta_{\rho\times\sigma}^{(k-1)}(B|X)   \right)=\zeta_{\rho\times\sigma}^{(k)}(B|x), 
\end{align}
where the second and third inequality hold due to Fatou's lemma and the induction hypothesis. Hence, for a given $\rho$ and randomized policies $\sigma(x)$, the unique stationary surplus distribution $\zeta_{\rho_n\times\sigma_n}^{(k)}(B|x)$ converges pointwise to $\zeta_{\rho\times\sigma}^{(k)}(B|x).$

Now by Lemma~\ref{eq:ksteptransitionkernel} and Equation~(\ref{eq:transitionrelation}), we need to show that $\liminf_{n\rightarrow\infty}\zeta_{\rho_n\times\sigma_n}(B)\geq \zeta_{\rho\times\sigma}(B)$. By Fatou's lemma, we have 
\begin{align}
&\liminf_{n\rightarrow\infty}\zeta_{\rho_n\times\sigma_n}(B)=\liminf_{n\rightarrow\infty}\sum_{k=0}^{\infty}(1-\beta){\beta}^k \mathbb{E}_{\Psi}\left(  \zeta_{\rho_n\times\sigma_n}^{(k)}(B|X_n)\right)\nonumber\displaybreak[0]\\
&\geq \sum_{k=0}^{\infty}(1-\beta){\beta}^k\mathbb{E}_{\Psi}\left( \liminf_{n\rightarrow\infty}  \zeta_{\rho_n\times\sigma_n}^{(k)}(B|X_n)  \right)\geq \sum_{k=0}^{\infty}(1-\beta){\beta}^k\mathbb{E}_{\Psi}\left(  \zeta_{\rho\times\sigma}^{(k)}(B|X)  \right)=\zeta_{\rho\times\sigma}(B).
\end{align}
Thus, for a given $\rho$ and the randomize policies $\sigma(x)$, the unique stationary surplus distribution $\zeta_{\rho_n\times\sigma_n}$ converges pointwise to $\zeta_{\rho\times\sigma}$. Then the stationary surplus distribution $\zeta_{\rho\times\sigma}$ is continuous in $\rho$ and $\sigma(x)$ for every surplus $x\in\mathbb{X}$.

{\bf{Proof of Lemma~\ref{claim1}}}\label{sec:Claim1}

Given the stationary surplus distribution $\zeta(x)$ and the action distribution $\sigma(x)$ at every surplus $x$,  those will introduce a population profile based on the actions chosen at each point $x$, denoted that action distribution as $\rho$, and we have $\rho_a=\sum_{x\in \mathbb{X}} \zeta(x)\cdot\sigma_a(x)$, where $a\in\mathcal{A}$, $\mathbb{X}$ is a countable set and $\mathcal{A}$ is a finite set. 

To show that $\rho$ is continuous in $\zeta(x)$ and $\sigma(x)$, we only need to show that for any sequence $\{\zeta_n\}_{n=1}^{\infty}$ converging to $\zeta$ in uniform norm, $\{\sigma_n(x)\}_{n=1}^{\infty}$ converging to $\sigma(x)$ pointwise,  we have $\{\rho_n\}_{n=1}^{\infty}$ converges to $\rho$ pointwise, which is equivalent to convergence in uniform norm as we have a finite set $\mathcal{A}$.

Since $\zeta_n\rightarrow\zeta$ uniformly, we have $\forall \epsilon_1>0$, $\exists N_1\in\mathbb{N},$ so that $\forall n\geq N_1$, $\forall x\in \mathbb{X}$, $|\zeta_n(x)-\zeta(x)|\leq\epsilon_1$. Similarly,   $\{\sigma_n(x)\}_{n=1}^{\infty}$ converges to $\sigma(x)$ pointwise, we have $\forall x\in \mathbb{X}$, and $\forall \epsilon_2>0$, $\exists N_2\in\mathbb{N}$ so that $\forall n\geq N_2$, , $|\sigma_n(x)-\sigma(x)|\leq\epsilon_2$. 
Now consider $\forall \epsilon=\max(\epsilon_1, \epsilon_2)$, we can find an all but finite subset $\mathbb{X}_1$ of $\mathbb{X}$, such that $\sum_{x\in\mathbb{X}_1}\zeta(x)\leq\frac{\epsilon}{2}$. Let $N=\max(N_1, N_2)$,  for $\forall x\in\mathbb{X}\setminus\mathbb{X}_1$, $\exists n > N$ large enough, such that $|\sigma_{n, a}(x)-\sigma_{a}(x)|\leq\frac{\epsilon}{2}$. Then $\forall x\in \mathbb{X}$, $\forall a\in\mathcal{A}$, we have
\begin{align}
&|\rho_{n, a}-\rho_a|=\left|\sum_x\zeta_n(x)\sigma_{n, a}(x)-\sum_x\zeta(x)\sigma_a(x)\right|\nonumber\\
                              &=\left|\sum_x\zeta_n(x)\sigma_{n, a}(x)-\sum_x\zeta_n(x)\sigma_a(x)+\sum_x\zeta_n(x)\sigma_a(x)-\sum_x\zeta(x)\sigma_a(x)\right|\nonumber\\
                              &\leq \left|\sum_x\zeta_n(x)\sigma_{n, a}(x)-\sum_x\zeta_n(x)\sigma_a(x)\right|+\left|\sum_x\zeta_n(x)\sigma_a(x)-\sum_x\zeta(x)\sigma_a(x)\right|\nonumber\\
                              &\leq \sum_x\zeta_n(x)|\sigma_{n, a}(x)-\sigma_a(x)|+\sum_x\sigma_a(x)|\zeta_n(x)-\zeta(x)|\nonumber\displaybreak[0]\\
                              &=\sum_{x\in\mathbb{X}_1}\zeta_n(x)|\sigma_{n, a}(x)-\sigma_a(x)|+\sum_{x\in\mathbb{X}\setminus\mathbb{X}_1}\zeta_n(x)|\sigma_{n, a}(x)-\sigma_a(x)|+\sum_x\sigma_a(x)|\zeta_n(x)-\zeta(x)|\nonumber\displaybreak[1]\\
                              &\stackrel{(a)}{\leq} \sum_{x\in\mathbb{X}_1}\zeta_n(x)\cdot 1+\sum_{x\in\mathbb{X}\setminus\mathbb{X}_1}\zeta_n(x)\cdot \frac{\epsilon}{2}+\sum_x\sigma_a(x)\cdot\epsilon_1\nonumber\displaybreak[2]\\
                              &\stackrel{(b)}{\leq} \frac{\epsilon}{2}\cdot 1+1\cdot\frac{\epsilon}{2}+\epsilon_1\cdot 1                              
                              \leq \epsilon\cdot 1+\epsilon\cdot 1
                              =2\epsilon,
\end{align}
where (a) follows from the fact that $|\sigma_{n, a}(x)-\sigma_{a}(x)|\leq1$ for $\forall x\in\mathbb{X}$, and $|\sigma_{n, a}(x)-\sigma_{a}(x)|<\frac{\epsilon}{2}$, for $x\in\mathbb{X}\setminus\mathbb{X}_1$ given $\epsilon>0$ and $n$ large enough, and the convergence of $\zeta_n$. (b) follows from that $\sum_{x\in\mathbb{X}_1}\zeta(x)<\frac{\epsilon}{2}$ for $x\in\mathbb{X}_1$. 

Therefore, $|\rho_{n, a}-\rho_a|<2\epsilon$ for all $a\in\mathcal{A}$ and $\forall n\geq N$, hence $\rho_n\rightarrow\rho$ pointwise, which is equivalent to convergence in uniform norm as we have a finite set $\mathcal{A}$.

\section{Characteristics of the best response policy}
{\bf{Proof of Lemma~\ref{lem:thresholdpolicy}}}\label{sec:Thresholdpolicy}

First, we consider $x\in\X$ and $x\geq 0$. We have
\begin{align}
&u(x)-\theta_{a_2(x)}+\beta [p_{\rho, a_2}(x)V_\rho(x+w)
+(1-p_{\rho, a_2}(x))V_\rho(x-l)]\nonumber\displaybreak[0] \\
&\gtrless u(x)-\theta_{a_1(x)}+\beta [p_{\rho, a_1}(x)V_\rho(x+w)
+(1-p_{\rho, a_1}(x))V_\rho(x-l)] \nonumber\displaybreak[1]\\
&\Leftrightarrow\theta_{a_1(x)}-\theta_{a_2(x)} \gtrless\beta[(p_{\rho, a_1}(x)-p_{\rho, a_2}(x))V_\rho(x+w)\nonumber\displaybreak[2]\\
&\qquad\qquad\qquad\qquad+((1-p_{\rho, a_1}(x))-(1-p_{\rho, a_2}(x)))V_\rho(x-l)]\nonumber\\
&\Leftrightarrow\theta_{a_1(x)}-\theta_{a_2(x)}\gtrless \beta(p_{\rho, a_1}(x)-p_{\rho, a_2}(x))[V_\rho(x+w)-V_\rho(x-l)].
\end{align}
As we assumed $\theta_{a_1(x)}>\theta_{a_2(x)}$, it follows that $p_{\rho, a_1}(x)>p_{\rho, a_2}(x)$. Also, since $w+l>0$ and $V_\rho(x)$ is increasing in $x$, so both sides of the above inequality are non-negative. Since $V_\rho(x)$ is submodular when $x\geq -l$, the RHS is a decreasing function of $x$.  Let $x_{a_1,a_2}^*\in\X$ be the smallest value such that $\text{LHS}\geq\text{RHS}$, then for all $x> x_{a_1,a_2}^*$ action $a_2(x)$ is preferred to action $a_1(x)$, for all $x < x_{a_1,a_2}^*$ action $a_1(x)$ is preferred to action $a_2(x)$, and finally, if at $x_{a_1,a_2}^*$ $\text{LHS=RHS}$, then at $x_{a_1,a_2}^*$ the agent is indifferent between the two actions, and if instead $\text{LHS} > \text{RHS}$, then action $a_2(x)$ is preferred to action $a_1(x)$.  We call $x_{a_1,a_2}^*$ the threshold value of surplus for actions $a_1(x)$ and $a_2(x)$.

Similarly, for $x\in\X$ and $x\leq 0$, $V_{\rho}(x)$ is supermodular when $x\leq w$, which implies the existence of a threshold policy.

{\bf{Proof of Lemma~\ref{Vcontinuous}}}\label{sec:Vsubmodular}

First, let $f\in \Phi$, suppose that $f$ is an increasing and submodular function.  First we prove that $T_\rho f$ is increasing and submodular too. Let $a^*(x)$ be an optimal action in the definition of $T_\rho f(x)$ when the surplus is $x$, i.e., one of the maximizers from \eqref{operator}. Let $x_1>x_2$, then 
\begin{align*}
&T_\rho f(x_1)-T_\rho f(x_2) = u(x_1)-u(x_2)-\theta_{a^*(x_1)}+\theta_{a^*(x_2)} +\beta \big[p_{\rho, a^*(x_1)}(x_1)f(x_1+w)+\nonumber\displaybreak[0]\\
&\qquad(1-p_{\rho, a^*(x_1)}(x_1))f(x_1-l)- p_{\rho, a^*(x_2)}(x_2)f(x_2+w)-(1-p_{\rho, a^*(x_2)}(x_2))f(x_2-l)\big] \nonumber\displaybreak[1]\\
& \geq u(x_1)-u(x_2)-\theta_{a^*(x_2)}+\theta_{a^*(x_2)}+\beta \big[p_{\rho, a^*(x_2)}(x_2)f(x_1+w)\nonumber\displaybreak[2]\\
&\qquad+(1-p_{\rho, a^*(x_2)}(x_2))f(x_1-l)- p_{\rho, a^*(x_2)}(x_2)f(x_2+w)-(1-p_{\rho, a^*(x_2)}(x_2))f(x_2-l)\big] \\
&=  u(x_1)-u(x_2)+\beta \big[ p_{\rho, a^*(x_2)}(x_2)(f(x_1+w+a)-f(x_2+w)\\
&\qquad+(1-p_{\rho, a^*(x_2)}(x_2))(f(x_1-l)-f(x_2-l)) \big]\geq 0.
\end{align*}
The first inequality holds because $a^*(x_2)$ need not be an optimal action when the surplus is $x_1$. 

Again, let $x_1>x_2$ and let $x>0$. Since $u(\cdot)$ is a concave function, it follows that it is submodular, i.e., 
\begin{align*}
 u(x_1+x)-u(x_1) \leq u(x_2+x) - u(x_2) 
\Leftrightarrow u(x_1+x) + u(x_2) \leq u(x_2+x) + u(x_1).
\end{align*}
Assuming that $f\in \Phi$ is submodular, we will now show that $T_\rho f$ is also submodular. 
Consider 
\begin{align*}
&T_\rho f(x_1+x) + T_\rho f(x_2) = u(x_1+x)+u(x_2)-\theta_{a^*(x_1+x)} - \theta_{a^*(x_2)}  \\
&+ \beta \big[ p_{\rho, a^*(x_1+x)}(x_1+x) f(x_1+x+w) + p_{\rho, a^*(x_2)}(x_2) f(x_2+w) \\
&+ (1-p_{\rho, a^*(x_1+x)}(x_1+x)) f(x_1+x-l) + (1-p_{\rho, a^*(x_2)}(x_2)) f(x_2-l) \big].
\end{align*}
We assume without loss of generality that $p_{\rho, a^*(x_1+x)}(x_1+x) \geq p_{\rho, a^*(x_2)}(x_2)$ and let $\delta$ be the difference; if $p_{\rho, a^*(x_1+x)}(x_1+x) \leq p_{\rho, a^*(x_2)}(x_2)$, then a similar proof establishes the result. Using this we have the RHS (denoted by $d$) being
\begin{align*}
& d=u(x_1+x)+u(x_2)-\theta_{a^*(x_1+x)} - \theta_{a^*(x_2)} +\beta \big[ p_{\rho, a^*(x_2)}(x_2) (f(x_1+x+w) + f(x_2+w))\\
& + (1-p_{\rho, a^*(x_1+x)}(x_1+x)) (f(x_1+x-l)+f(x_2-l)) + \delta (f(x_1+x+w)+f(x_2-l) ) \big].
\end{align*}
By submodularity of $f(\cdot)$ we have
\begin{align*}
f(x_1+x+w) + f(x_2+w) \leq f(x_2+x+w)+f(x_1+w), \\
f(x_1+x-l)+f(x_2-l) \leq f(x_2+x-l)+f(x_1-l), \\
f(x_1+x+w)+f(x_2-l) \leq f(x_2+x+w)+f(x_1-l).
\end{align*}
With these and using the submodularity of $u(\cdot)$ we get
\begin{align}
& d \leq u(x_2+x)+u(x_1)-\theta_{a^*(x_1+x)} - \theta_{a^*(x_2)}  +\beta \big[ p_{\rho, a^*(x_2)}(x_2) (f(x_2+x+w) + f(x_1+w))\nonumber\displaybreak[0]\\
& \quad + (1-p_{\rho, a^*(x_1+x)}(x_1+x)) (f(x_2+x-l)+f(x_1-l)) + \delta (f(x_2+x+w)+f(x_1-l) ) \big] \nonumber\displaybreak[1]\\
& = u(x_2+x) -\theta_{a^*(x_1+x)} + \beta[ p_{\rho, a^*(x_2)}(x_2) f(x_2+x+w)  + (1-p_{\rho, a^*(x_2)}(x_2)) f(x_2+x-l)]\nonumber\displaybreak[2] \\
& \quad + u(x_1) -\theta_{a^*(x_2)} + \beta[ p_{\rho, a^*(x_2)}(x_2) f(x_1+w) + (1-p_{\rho, a^*(x_1+x)}(x_1+x)) f(x_1-l)]\nonumber\displaybreak[3] \\
& \leq T_\rho f(x_2+x) + T_\rho f(x_1),\nonumber
\end{align}
where the last inequality holds as using the optimal actions $(a^*(x_2+x),a^*(x_1))$ yields a higher value as opposed to the sub-optimal actions $(a^*(x_1+x),a^*(x_2))$ when the surplus is $x_2+x$ and $x_1$.

Since both the monotonicity and submodularity properties are preserved when taking pointwise limits, choosing $f(\cdot)\equiv 0$ (or $u(\cdot)$) to start the value iteration proves that the value function $V_\rho(\cdot)$ is increasing and submodular.

Similarly, if $f\in\Phi$ is an increasing and supermodular function, following the same argument, we can prove that the value function $V_{\rho}(\cdot)$ is increasing and supermodular.

\bibliographystyle{ACM-Reference-Format-Journals}
\bibliography{refs}


\begin{thebibliography}{00}


\ifx \showCODEN    \undefined \def \showCODEN     #1{\unskip}     \fi
\ifx \showDOI      \undefined \def \showDOI       #1{{\tt DOI:}\penalty0{#1}\ }
  \fi
\ifx \showISBNx    \undefined \def \showISBNx     #1{\unskip}     \fi
\ifx \showISBNxiii \undefined \def \showISBNxiii  #1{\unskip}     \fi
\ifx \showISSN     \undefined \def \showISSN      #1{\unskip}     \fi
\ifx \showLCCN     \undefined \def \showLCCN      #1{\unskip}     \fi
\ifx \shownote     \undefined \def \shownote      #1{#1}          \fi
\ifx \showarticletitle \undefined \def \showarticletitle #1{#1}   \fi
\ifx \showURL      \undefined \def \showURL       #1{#1}          \fi

\bibitem[\protect\citeauthoryear{Adlakha, Johari, and Weintraub}{Adlakha
  et~al\mbox{.}}{2015}]%
        {adlakha15}
{S. Adlakha}, {R. Johari}, {and} {G. Weintraub}. 2015.
\newblock \showarticletitle{Equilibria of dynamic games with many players:
  Existence, approximation, and market structure}.
\newblock {\em Journal of Economic Theory\/}  {156} (2015), 269--316.
\newblock


\bibitem[\protect\citeauthoryear{Albadi and El-Saadany}{Albadi and
  El-Saadany}{2008}]%
        {AlbSaa08}
{M.~H. Albadi} {and} {E.~F. El-Saadany}. 2008.
\newblock \showarticletitle{A summary of demand response in electricity
  markets}.
\newblock {\em Electric Power Systems Research\/} {78}, 11 (2008), 1989--1996.
\newblock


\bibitem[\protect\citeauthoryear{Allcott and Kessler}{Allcott and
  Kessler}{2015}]%
        {allcott2015welfare}
{H. Allcott} {and} {J. Kessler}. 2015.
\newblock {\em The Welfare Effects of Nudges: A Case Study of Energy Use Social
  Comparisons}.
\newblock {T}echnical {R}eport. National Bureau of Economic Research.
\newblock


\bibitem[\protect\citeauthoryear{Billingsley}{Billingsley}{2013}]%
        {billingsley2013}
{P. Billingsley}. 2013.
\newblock {\em Convergence of probability measures}.
\newblock John Wiley \& Sons.
\newblock


\bibitem[\protect\citeauthoryear{Bitar}{Bitar}{2015}]%
        {bitar15}
{E. Bitar}. 2015.
\newblock Coordinated Aggregation of Distributed Demand-Side Resources.
\newblock   (2015).
\newblock
\newblock
\shownote{\url{http://www.news.cornell.edu/stories/2015/03/adding-renewable-energy-power-grid-requires-flexibility}.}


\bibitem[\protect\citeauthoryear{Borkar and Sundaresan}{Borkar and
  Sundaresan}{2013}]%
        {borkar13}
{Vivek~S Borkar} {and} {Rajesh Sundaresan}. 2013.
\newblock \showarticletitle{Asymptotics of the {I}nvariant {M}easure in {M}ean
  {F}ield {M}odels with {J}umps}.
\newblock {\em Stochastic Systems\/} {2}, 2 (2013), 322--380.
\newblock


\bibitem[\protect\citeauthoryear{Callaway}{Callaway}{2009}]%
        {Cal09}
{D.~S. Callaway}. 2009.
\newblock \showarticletitle{Tapping the energy storage potential in electric
  loads to deliver load following and regulation, with application to wind
  energy}.
\newblock {\em Energy Conversion and Management\/} (2009).
\newblock


\bibitem[\protect\citeauthoryear{Chen and Wang}{Chen and Wang}{2010}]%
        {chen2010tax}
{S. Chen} {and} {J. Wang}. 2010.
\newblock \showarticletitle{Tax Evasion and Fraud Detection: A Theoretical
  Evaluation of {T}aiwan's Business Tax Policy for Internet Auctions}.
\newblock {\em Asian Social Science\/} {6}, 12 (2010), 23.
\newblock


\bibitem[\protect\citeauthoryear{Clark, Wroclawski, Sollins, and Braden}{Clark
  et~al\mbox{.}}{2002}]%
        {clark2002tussle}
{D.~D. Clark}, {J. Wroclawski}, {K.~R. Sollins}, {and} {R. Braden}. 2002.
\newblock \showarticletitle{Tussle in cyberspace: defining tomorrow's
  {I}nternet}.
\newblock {\em ACM SIGCOMM Computer Communication Review\/} {32}, 4 (2002),
  347--356.
\newblock


\bibitem[\protect\citeauthoryear{ERCOT}{ERCOT}{2014}]%
        {ercot}
{ERCOT}. 2014.
\newblock {E}lectric {R}eliability {C}ouncil of {T}exas ({ERCOT}).
\newblock   (2014).
\newblock
\newblock
\shownote{Data set available at \url{http://www.ercot.com/}.}


\bibitem[\protect\citeauthoryear{Gao, Frejinger, and Ben-Akiva}{Gao
  et~al\mbox{.}}{2010}]%
        {gao2010adaptive}
{S. Gao}, {E. Frejinger}, {and} {M. Ben-Akiva}. 2010.
\newblock \showarticletitle{Adaptive route choices in risky traffic networks: A
  prospect theory approach}.
\newblock {\em Transportation research part C: emerging technologies\/} {18}, 5
  (2010), 727--740.
\newblock


\bibitem[\protect\citeauthoryear{Gomes, Mohr, and Souza}{Gomes
  et~al\mbox{.}}{2010}]%
        {gomes10}
{D. Gomes}, {J. Mohr}, {and} {R.R. Souza}. 2010.
\newblock \showarticletitle{Discrete time, finite state space mean field
  games}.
\newblock {\em Journal de math{\'e}matiques pures et appliqu{\'e}es\/} {93}, 3
  (2010), 308--328.
\newblock


\bibitem[\protect\citeauthoryear{Graham and M{\'e}l{\'e}ard}{Graham and
  M{\'e}l{\'e}ard}{1994}]%
        {GraMel94}
{C. Graham} {and} {S. M{\'e}l{\'e}ard}. 1994.
\newblock \showarticletitle{Chaos hypothesis for a system interacting through
  shared resources}.
\newblock {\em Probability Theory and Related Fields\/} {100}, 2 (1994),
  157--174.
\newblock


\bibitem[\protect\citeauthoryear{Hao, Sanandaji, Poolla, and Vincent}{Hao
  et~al\mbox{.}}{2015}]%
        {Hao14}
{H. Hao}, {B.~M. Sanandaji}, {K. Poolla}, {and} {T.~L. Vincent}. 2015.
\newblock \showarticletitle{Aggregate Flexibility of Thermostatically
  Controlled Loads}.
\newblock {\em IEEE Transactions on Power Systems\/} {30}, 1 (2015), 189--198.
\newblock


\bibitem[\protect\citeauthoryear{Hao and Xie}{Hao and Xie}{2014}]%
        {HaoXie14}
{M. Hao} {and} {L. Xie}. 2014.
\newblock \showarticletitle{Analysis of Coupon Incentive-Based Demand Response
  with Bounded Consumer Rationality}. In {\em North American Power Symposium}.
  1--6.
\newblock


\bibitem[\protect\citeauthoryear{Harrison and Rutstr{\"o}m}{Harrison and
  Rutstr{\"o}m}{2009}]%
        {harrison2009expected}
{G.~W. Harrison} {and} {E.~E. Rutstr{\"o}m}. 2009.
\newblock \showarticletitle{Expected utility theory and prospect theory: One
  wedding and a decent funeral}.
\newblock {\em Experimental Economics\/} {12}, 2 (2009), 133--158.
\newblock


\bibitem[\protect\citeauthoryear{Huang, Malham{\'e}, and Caines}{Huang
  et~al\mbox{.}}{2006}]%
        {huang2006large}
{M. Huang}, {R.~P. Malham{\'e}}, {and} {P.~E. Caines}. 2006.
\newblock \showarticletitle{Large population stochastic dynamic games:
  closed-loop {McKean-Vlasov} systems and the {N}ash certainty equivalence
  principle}.
\newblock {\em Communications in Information \& Systems\/} {6}, 3 (2006),
  221--252.
\newblock


\bibitem[\protect\citeauthoryear{Hunter}{Hunter}{2004}]%
        {hunter2004mm}
{D.~R. Hunter}. 2004.
\newblock \showarticletitle{{MM} algorithms for generalized Bradley-Terry
  models}.
\newblock {\em Annals of Statistics\/} (2004).
\newblock


\bibitem[\protect\citeauthoryear{Iyer, Johari, and Sundararajan}{Iyer
  et~al\mbox{.}}{2014}]%
        {IyeJoh14}
{K. Iyer}, {R. Johari}, {and} {M. Sundararajan}. 2014.
\newblock \showarticletitle{Mean field equilibria of dynamic auctions with
  learning}.
\newblock {\em Management Science\/} {60}, 12 (2014), 2949--2970.
\newblock


\bibitem[\protect\citeauthoryear{Jovanovic and Rosenthal}{Jovanovic and
  Rosenthal}{1988}]%
        {Boyan88}
{B. Jovanovic} {and} {R.~W. Rosenthal}. 1988.
\newblock \showarticletitle{{Anonymous sequential games}}.
\newblock {\em Journal of Mathematical Economics\/} {17}, 1 (February 1988),
  77--87.
\newblock


\bibitem[\protect\citeauthoryear{Kahneman and Tversky}{Kahneman and
  Tversky}{1979}]%
        {kahneman1979prospect}
{D. Kahneman} {and} {A. Tversky}. 1979.
\newblock \showarticletitle{Prospect theory: An analysis of decision under
  risk}.
\newblock {\em Econometrica: Journal of the Econometric Society\/} (1979),
  263--291.
\newblock


\bibitem[\protect\citeauthoryear{Kahneman and Tversky}{Kahneman and
  Tversky}{1984}]%
        {kahneman1984choices}
{D. Kahneman} {and} {A. Tversky}. 1984.
\newblock \showarticletitle{Choices, values, and frames.}
\newblock {\em American psychologist\/} {39}, 4 (1984), 341.
\newblock


\bibitem[\protect\citeauthoryear{Lasry and Lions}{Lasry and Lions}{2007}]%
        {LasLio07}
{J.~M. Lasry} {and} {P.~L. Lions}. 2007.
\newblock \showarticletitle{Mean field games}.
\newblock {\em Japanese Journal of Mathematics\/} {2}, 1 (2007), 229--260.
\newblock


\bibitem[\protect\citeauthoryear{Li, Bhattacharyya, Paul, Shakkottai, and
  Subramanian}{Li et~al\mbox{.}}{2016a}]%
        {JianRaj16ton}
{J. Li}, {R. Bhattacharyya}, {S. Paul}, {S. Shakkottai}, {and} {V.
  Subramanian}. 2016a.
\newblock \showarticletitle{Incentivizing Sharing in Realtime {D2D} Streaming
  Networks: A Mean Field Game Perspective}.
\newblock {\em IEEE/ACM Transactions on Networking\/} (2016).
\newblock


\bibitem[\protect\citeauthoryear{Li, Bhattacharyya, Paul, Shakkottai, and
  Subramanian}{Li et~al\mbox{.}}{2016b}]%
        {LiBha15arxiv}
{J. Li}, {R. Bhattacharyya}, {S. Paul}, {S. Shakkottai}, {and} {V.
  Subramanian}. 2016b.
\newblock \showarticletitle{Incentivizing {S}haring in {R}ealtime {D2D}
  {S}treaming {N}etworks: {A} {M}ean {F}ield {G}ame {P}erspective}.
\newblock {\em arXiv preprint arXiv:1604.02435\/} (2016).
\newblock


\bibitem[\protect\citeauthoryear{Li, Xia, Geng, Hao, Shakkottai, Subramanian,
  and Le}{Li et~al\mbox{.}}{2015}]%
        {JianBai15}
{J. Li}, {B. Xia}, {X. Geng}, {M. Hao}, {S. Shakkottai}, {V. Subramanian},
  {and} {X. Le}. 2015.
\newblock \showarticletitle{Energy Coupon: A Mean Field Game Perspective on
  Demand Response in Smart Grids}. In {\em Proceedings of ACM SIGMETRICS}.
  455--456.
\newblock


\bibitem[\protect\citeauthoryear{Li and Mandayam}{Li and Mandayam}{2012}]%
        {li2012prospects}
{T. Li} {and} {N.~B. Mandayam}. 2012.
\newblock \showarticletitle{Prospects in a wireless random access game}. In
  {\em Proceedings of Conference on Information Sciences and Systems (CISS)}.
  1--6.
\newblock


\bibitem[\protect\citeauthoryear{Loiseau, Schwartz, Musacchio, Amin, and
  Sastry}{Loiseau et~al\mbox{.}}{2014}]%
        {patrick14}
{P. Loiseau}, {G.~A. Schwartz}, {J. Musacchio}, {S. Amin}, {and} {S.~S.
  Sastry}. 2014.
\newblock \showarticletitle{Incentive Mechanisms for {I}nternet Congestion
  Management: Fixed-Budget Rebate Versus Time-of-Day Pricing}.
\newblock {\em IEEE/ACM Transactions on Networking\/} {22}, 2 (April 2014),
  647--661.
\newblock


\bibitem[\protect\citeauthoryear{Lozano and Irurozki}{Lozano and
  Irurozki}{2012}]%
        {lozano12}
{J.~A. Lozano} {and} {E. Irurozki}. 2012.
\newblock Probabilistic modeling on rankings.
\newblock   (2012).
\newblock
\newblock
\shownote{available at
  \url{http://www.sc.ehu.es/ccwbayes/members/ekhine/tutorial_ranking/info.html}.}


\bibitem[\protect\citeauthoryear{Manjrekar, Ramaswamy, and
  Shakkottai}{Manjrekar et~al\mbox{.}}{2014}]%
        {ManRam14}
{M. Manjrekar}, {V. Ramaswamy}, {and} {S. Shakkottai}. 2014.
\newblock \showarticletitle{A Mean Field Game Approach to Scheduling in
  Cellular Systems}. In {\em Proceedings of IEEE Infocom}. Toronto, Canada.
\newblock


\bibitem[\protect\citeauthoryear{Merugu, Prabhakar, and Rama}{Merugu
  et~al\mbox{.}}{2009}]%
        {MerPra09}
{D. Merugu}, {B.~S. Prabhakar}, {and} {N.~S. Rama}. 2009.
\newblock \showarticletitle{An Incentive Mechanism for Decongesting the Roads:
  A Pilot Program in {B}angalore}. In {\em Proceedings of NetEcon, ACM Workshop
  on the Economics of Networked Systems}.
\newblock


\bibitem[\protect\citeauthoryear{Meyn and Tweedie}{Meyn and Tweedie}{2009}]%
        {MeyTwe09}
{S.~P. Meyn} {and} {R.~L. Tweedie}. 2009.
\newblock {\em {Markov chains and stochastic stability}}.
\newblock Cambridge University Press.
\newblock


\bibitem[\protect\citeauthoryear{Naritomi}{Naritomi}{2013}]%
        {jnari13}
{J. Naritomi}. 2013.
\newblock {\em {Consumers as Tax Auditors}}.
\newblock working paper, International Development Department and Institute of
  Public Affairs, London School of Economics.
\newblock


\bibitem[\protect\citeauthoryear{OhmConnect}{OhmConnect}{2015}]%
        {ohm-connect}
{OhmConnect}. 2015.
\newblock   (2015).
\newblock
\newblock
\shownote{Online at \url{https://www.ohmconnect.com/}.}


\bibitem[\protect\citeauthoryear{Pecan-Street}{Pecan-Street}{2014}]%
        {pecanstreet}
{Pecan-Street}. 2014.
\newblock   (2014).
\newblock
\newblock
\shownote{Data set available at \url{https://dataport.pecanstreet.org/}.}


\bibitem[\protect\citeauthoryear{Poco, Lopes, and Silva}{Poco
  et~al\mbox{.}}{2015}]%
        {polo15}
{M. Poco}, {C. Lopes}, {and} {A. Silva}. 2015.
\newblock {\em {Perception of Tax Evasion and Tax Fraud in Portugal: A
  Sociological Study}}.
\newblock working paper.
\newblock


\bibitem[\protect\citeauthoryear{Prabhakar}{Prabhakar}{2013}]%
        {Pra13}
{B. Prabhakar}. 2013.
\newblock \showarticletitle{Designing Large-scale Nudge Engines}. In {\em
  Proceedings of the ACM SIGMETRICS}. 1--2.
\newblock


\bibitem[\protect\citeauthoryear{Prelec}{Prelec}{1998}]%
        {prelec1998probability}
{D. Prelec}. 1998.
\newblock \showarticletitle{The probability weighting function}.
\newblock {\em Econometrica\/} (1998), 497--527.
\newblock


\bibitem[\protect\citeauthoryear{Puterman}{Puterman}{1994}]%
        {Put94}
{M.~L. Puterman}. 1994.
\newblock {\em {Markov decision processes: Discrete stochastic dynamic
  programming}}.
\newblock John Wiley \& Sons, Inc.
\newblock


\bibitem[\protect\citeauthoryear{Qin, Geng, and Liu}{Qin et~al\mbox{.}}{2010}]%
        {qin2010new}
{T. Qin}, {X. Geng}, {and} {T. Liu}. 2010.
\newblock \showarticletitle{A new probabilistic model for rank aggregation}. In
  {\em Advances in neural information processing systems}. 1948--1956.
\newblock


\bibitem[\protect\citeauthoryear{Ross}{Ross}{2013}]%
        {ross2013applied}
{Sheldon~M Ross}. 2013.
\newblock {\em Applied {P}robability {M}odels with {O}ptimization
  {A}pplications}.
\newblock Courier Corporation.
\newblock


\bibitem[\protect\citeauthoryear{Schwartz, Tembine, Amin, and Sastry}{Schwartz
  et~al\mbox{.}}{2012}]%
        {galina12}
{G.~A. Schwartz}, {H. Tembine}, {S. Amin}, {and} {S.~S. Sastry}. 2012.
\newblock \showarticletitle{Electricity Demand Shaping via Randomized Rewards:
  A Mean Field Game Approach}.
\newblock {\em Allerton Conference on Communication, Control, and Computing\/}
  (2012).
\newblock


\bibitem[\protect\citeauthoryear{Tversky and Kahneman}{Tversky and
  Kahneman}{1981}]%
        {tversky1981framing}
{A. Tversky} {and} {D. Kahneman}. 1981.
\newblock \showarticletitle{The framing of decisions and the psychology of
  choice}.
\newblock {\em Science\/} {211}, 4481 (1981), 453--458.
\newblock


\bibitem[\protect\citeauthoryear{Tversky and Kahneman}{Tversky and
  Kahneman}{1992}]%
        {tversky1992advances}
{A. Tversky} {and} {D. Kahneman}. 1992.
\newblock \showarticletitle{Advances in prospect theory: Cumulative
  representation of uncertainty}.
\newblock {\em Journal of Risk and uncertainty\/} {5}, 4 (1992), 297--323.
\newblock


\bibitem[\protect\citeauthoryear{Wang, Saad, Mandayam, and Poor}{Wang
  et~al\mbox{.}}{2014}]%
        {wang2014integrating}
{Y. Wang}, {W. Saad}, {N.~B. Mandayam}, {and} {H.~V. Poor}. 2014.
\newblock \showarticletitle{Integrating energy storage into the smart grid: A
  prospect theoretic approach}. In {\em Proceedings of Acoustics, Speech and
  Signal Processing (ICASSP)}. 7779--7783.
\newblock


\bibitem[\protect\citeauthoryear{Xiao, Mandayam, and Poor}{Xiao
  et~al\mbox{.}}{2015}]%
        {xiaoprospect}
{L. Xiao}, {N.~B. Mandayam}, {and} {H.~V. Poor}. 2015.
\newblock \showarticletitle{Prospect Theoretic Analysis of Energy Exchange
  Among Microgrids}.
\newblock {\em IEEE Transactions on Smart Grid\/} {6}, 1 (January 2015),
  63--72.
\newblock


\bibitem[\protect\citeauthoryear{Yu, Cheung, and Huang}{Yu
  et~al\mbox{.}}{2014}]%
        {yu2014spectrum}
{J. Yu}, {M.~H. Cheung}, {and} {J. Huang}. 2014.
\newblock \showarticletitle{Spectrum investment with uncertainty based on
  prospect theory}. In {\em Proceedings of International conference on
  Communications (ICC)}. 1620--1625.
\newblock


\bibitem[\protect\citeauthoryear{Zhong, Xie, and Xia}{Zhong
  et~al\mbox{.}}{2013}]%
        {ZhoXie13}
{H. Zhong}, {L. Xie}, {and} {Q. Xia}. 2013.
\newblock \showarticletitle{Coupon Incentive-Based Demand Response: Theory and
  Case Study}.
\newblock {\em IEEE Transactions on Power Systems\/} {28}, 2 (May 2013),
  1266--1276.
\newblock


\end{thebibliography}

\end{document}